\documentclass[12pt]{article}

\usepackage{latexsym}
\usepackage{amssymb,amsfonts,amsmath}
\usepackage{graphicx} 
\usepackage{indentfirst}
\usepackage{bbm}
\usepackage{amssymb}
\usepackage{verbatim}
\usepackage{amsmath, amsthm,amssymb}
\usepackage{mathrsfs}
\usepackage{hyperref}
\usepackage{amsfonts}
\usepackage{dsfont}
\usepackage{cite}
\usepackage{xcolor}
\usepackage[multiple]{footmisc}
\parskip=\medskipamount

\newcommand {\cD}{{\cal D}}
\newcommand {\cE}{{\cal E}}

\newcommand {\cK}{{\cal K}}

\newcommand {\cM}{{\cal M}}
\newcommand {\cN}{{\cal N}}
\newcommand {\cO}{{\cal O}}

\newcommand {\cS}{{\cal S}}

\def\a{\alpha}
\def \bi{\bibitem}

\def\b{\beta}

\def\d{\delta}

\def\f{\phi}
\def\g{\gamma}
\def\G{\Gamma}

\def\k{\kappa}
\def\l{\lambda}

\def\n{\nu}
\def\o{\omega}
\def\p{\pi}
\def\q{\theta}

\def\s{\sigma}

\def\x{\xi}

\def\D{\Delta}
\def\F{\Phi}

\def\L{\Lambda}
\def\O{\Omega}

\def\U{\Upsilon}
\def\X{\Xi}

\def\rd{{\rm d}}
\def\ri{{\rm i}}
\def\re{{\rm e}}

\newcommand{\ad}{{\dot{\alpha}}}                           %new
\newcommand{\bd}{{\dot{\beta}}}                            %new
\newcommand{\ve}{\varepsilon}                            %new
\newcommand{\cDB}{{\bar\cD}}                            %new

\newcommand{\pa}{\partial}                           %new
\newcommand{\hf}{\frac12}
\newcommand{\vf}{\varphi}
\newcommand{\be}{\begin{equation}}
\newcommand{\ee}{\end{equation}}
\newcommand{\bea}{\begin{eqnarray}}
\newcommand{\eea}{\end{eqnarray}}
\newcommand{\non}{\nonumber}
\newcommand{\1}{{\underline{1}}}
\newcommand{\2}{{\underline{2}}}

\newcommand{\dsC}{{\mathbb C}}

\newcommand{\bm}[1]{\mbox{\boldmath$#1$}}

\def\double #1{#1{\hbox{\kern-2pt $#1$}}}

\newcommand{\gd}{{\dot\g}}
\newcommand{\dd}{{\dot\d}}

\newcommand{\sba}{{\bar{\s}}}

\newcommand{\dalpha}{{\dot{\alpha}}}
\newcommand{\dbeta}{{\dot{\beta}}}

\newif\ifdtup

\newcommand{\bsubeq}{\begin{subequations}}
\newcommand{\esubeq}{\end{subequations}}
\newcommand{\fF}{\mathfrak F}

\numberwithin{equation}{section}

\newcommand{\sSU}{\mathsf{SU}}
\newcommand{\sSL}{\mathsf{SL}}

\newcommand{\sU}{\mathsf{U}}

\begin{document}

\begin{titlepage}
\begin{flushright}
June, 2021 \\
\end{flushright}
\vspace{5mm}

\begin{center}
{\Large \bf 
Superconformal duality-invariant models 
and $\mathcal{N} = 4$ SYM effective action} 
\end{center}

\begin{center}

{\bf Sergei M. Kuzenko} \\
\vspace{5mm}

\footnotesize{
{\it Department of Physics M013, The University of Western Australia\\
35 Stirling Highway, Perth W.A. 6009, Australia}}  
~\\
\vspace{2mm}
~\\
Email: \texttt{ 
sergei.kuzenko@uwa.edu.au}\\
\vspace{2mm}

\end{center}

\begin{abstract}
\baselineskip=14pt
We present $\mathcal{N}=2$ superconformal $\mathsf{U}(1)$ duality-invariant models for an Abelian vector multiplet coupled to conformal supergravity. In a Minkowski background, such a nonlinear theory is expected to describe (the planar part of) the low-energy effective action for the $\mathcal{N}=4$ $\mathsf{SU}(N)$ super-Yang-Mills (SYM) theory on its Coulomb branch where (i) the gauge group $\mathsf{SU}(N)$ is spontaneously broken to $\mathsf{SU}(N-1) \times \mathsf{U}(1)$; and (ii) the dynamics is captured by a single $\mathcal{N}=2$ vector multiplet associated with the $\mathsf{U}(1)$ factor of the unbroken group. Additionally, a local $\mathsf{U}(1)$ duality-invariant 
action generating the $\mathcal{N}=2$ super-Weyl anomaly is proposed. 
By providing a new derivation of the recently constructed $\mathsf{U}(1)$ duality-invariant $\mathcal{N}=1$ superconformal electrodynamics, we introduce its 
$\mathsf{SL}(2,{\mathbb R})$ duality-invariant coupling to the dilaton-axion multiplet.
\end{abstract}
\vspace{5mm}

\vfill

\vfill
\end{titlepage}

\newpage
\renewcommand{\thefootnote}{\arabic{footnote}}
\setcounter{footnote}{0}

\tableofcontents{}
\vspace{1cm}
\bigskip\hrule

\allowdisplaybreaks

%%%%%%%%%%%%%%%%%%%%%%%%%%%%%%%%
%%%%%%%%%%%%%%%%%%%%%%%%%%%%%%%%

\section{Introduction}

It is believed that the $\cN=4$ super Yang-Mills (SYM) theory is self-dual 
\cite{MO,Osborn} (see also \cite{Sen}). This conjecture was originally put forward in the late 1970s as a duality between the conventional and soliton sectors  of the theory. 
Twenty years later it was suggested\footnote{This was inspired in part by the Seiberg-Witten theory \cite{SW1,SW2} and also by the AdS/CFT correspondence 
\cite{Maldacena}.}
 \cite{GKPR}
that self-duality might be realised in terms of a low-energy 
effective action of the theory on its Coulomb branch. 
Here the gauge group $\sSU(N)$ is spontaneously 
broken to $\sSU(N-1) \times \sU(1)$ and the dynamics 
is described by a single $\cN=2$ vector multiplet associated with the $\sU(1)$ factor 
of the unbroken group. 
Two different realisations of self-duality for the $\cN=4$ SYM effective action
in $\cN=2$ superspace were  proposed:
(i) self-duality under Legendre transformation
\cite{GKPR}; and (ii) self-duality under 
$\sU(1)$ duality rotations \cite{KT1}.\footnote{It was also conjectured by Schwarz 
\cite{Schwarz:2013wra} that the world-volume action of a probe D3-brane in an 
$\rm AdS_5 \times S^5$ background of type IIB superstring theory, with one unit of flux, can be reinterpreted as the exact (or highly) effective action for 
$\sU(2)$ $\cN = 4$ super Yang-Mills theory on the Coulomb branch.
}
So far, both proposals have not been derived from first principles, 
although each of them is consistent with the 
one-loop \cite{PvU, G-RR, BK98, BBK,LvU} and
two-loop \cite{K2004} calculations.\footnote{For an alternative two-loop calculation see \cite{BPT}.}
It is worth pointing out that (ii) implies (i), see \cite{KT1,KT2} for the technical details. 

Building on the influential 1981 work by Gaillard and Zumino \cite{GZ1},
the general theory of $\sU(1)$ duality-invariant
models for nonlinear electrodynamics in four dimensions was developed in the mid 1990s \cite{GR1,GR2,GZ2,GZ3} and the early 2000s \cite{IZ_N3,IZ1,IZ2} (see also \cite{IZ3}), 
including the case of duality-invariant theories 
with higher derivatives \cite{KT2}.\footnote{Further aspects of duality-invariant theories 
with higher derivatives were studied, e.g., in \cite{AFZ,Chemissany:2011yv,AF,AFT}.} 
The Gaillard-Zumino-Gibbons-Rasheed  
(GZGR) formalism \cite{GZ1,GR1,GR2,GZ2,GZ3} admits a natural extension to higher dimensions \cite{Tanii,AT,ABBZ} (see also \cite{KT2,AFZ,Tanii2} for a review). 
In four dimensions, this setting has been properly generalised to formulate general 
$\sU(1)$ duality-invariant $\cN=1$ and $\cN=2$ globally \cite{KT1,KT2} and locally \cite{KMcC,KMcC2,K12} supersymmetric theories. 
In particular, extending the earlier proposal of \cite{Ketov}, the first consistent 
perturbative scheme to construct the $\cN=2$ supersymmetric Born-Infeld action 
was given in \cite{KT2}. The formalism of nonlinear realisations for
the partial  $\cN=4 \to \cN=2$ breaking of supersymmetry advocated in \cite{BIK1}
reproduced \cite{BIK2} the results of \cite{KT2}.
Further progress toward the construction of the $\cN=2$ supersymmetric Born-Infeld action has been achieved in \cite{BCFKR,CK,IZ4}. 

Eight years ago, the general formalism of supersymmetric duality rotations 
\cite{KT1,KT2,KMcC,KMcC2,K12} was combined with the bosonic approach due to 
Ivanov and Zupnik\footnote{The Ivanov-Zupnik (IZ) approach 
was inspired by the structure of the $\cN=3$ supersymmetric Born-Infeld action
proposed in \cite{IZ_N3}.}
 \cite{IZ1,IZ2,IZ3} to develop new formulations for $\cN=1$ and 
$\cN=2$ supersymmetric duality-invariant theories coupled to supergravity 
\cite{K13}. The method makes use of an auxiliary unconstrained chiral superfield (a spinor in the $\cN=1$ case and a scalar for $\cN=2$) and is characterised by the  fundamental property that $\sU(1)$ duality invariance is equivalent to the manifest $\sU(1)$ invariance of the self-interaction. In the $\cN=1$ rigid supersymmetric case,
analogous results were independently obtained in \cite{ILZ}.\footnote{The IZ approach  \cite{IZ1,IZ2}
has also been generalised to higher dimensions in \cite{K19duality}.
It has been used \cite{Novotny} to establish the relation between helicity conservation for the tree-level scattering amplitudes and the electric-magnetic duality.
In four dimensions a hybrid formulation has been developed \cite{INZ} which combines the powerful features of the IZ approach with the Pasti-Sorokin-Tonin formalism \cite{PST1,PST2}.
} 

A constructive perturbative scheme to compute $\cN=2$ superconformal 
$\sU(1)$ duality-invariant actions for the $\cN=2$ vector multiplet was described in \cite{KT1}. In the present paper we will give closed-form expressions for such actions using the formalism 
of Ref. \cite{K13} in conjunction with the results of earlier publications \cite{BdeWKL,K13anomaly} devoted to the $\cN=2$ super-Gauss-Bonnet term and
super-Weyl anomalies. 

While Ref. \cite{KT1} described the $\cN=2$ superconformal $\sU(1)$ duality-invariant actions, it did not discuss $\cN=0$ and $\cN=1 $ duality-invariant models compatible with (super) conformal symmetry. The consideration in \cite{KT1,KT2} was restricted to those
$\cN=0$ and $\cN=1 $ self-dual systems which are well-defined in a weak-field limit.\footnote{This means that the interactions $\L(z,\bar z)$ in \eqref{2.1} and \eqref{A.2}  were chosen in \cite{KT1,KT2} to be real analytic. As a result, (super)conformal nonlinear systems were automatically excluded.}
A year ago, it was shown that there exists a unique
model for conformal duality-invariant electrodynamics  \cite{BLST} (see also \cite{Kosyakov}). Recently its $\cN=1$ superconformal extension has been introduced 
\cite{BLST2}. Below we will show how the  $\mathcal{N}=1$ superconformal duality-invariant model of \cite{BLST2} naturally occurs within the framework developed in 
\cite{KT1,KT2,K13}.

This paper is organised as follows. As a warmup exercise, in section \ref{Section2} we give a new derivation of the recently constructed $\mathcal{N}=1$ superconformal duality-invariant model \cite{BLST2}. Section  \ref{Section3} describes the new
$\mathcal{N}=2$ superconformal duality-invariant models. 
The obtained results and some generalisations are discussed in section \ref{Section4}. 
The main body of the paper is accompanied by three technical appendices. Appendix \ref{AppendixA}  reviews the model for conformal duality-invariant electrodynamics  \cite{BLST} (see also \cite{Kosyakov}). 
Appendix \ref{AppendixB}  collects those results concerning $\cN=1$ supergravity and super-Weyl transformations, which are used in section 2. Appendix \ref{AppendixC}  contains  similar material but for the $\cN=2$ case. Our two-component spinor notation and conventions correspond to \cite{WB,Ideas}.

%%%%%%%%%%%%%%%%%%%%%%%%%%%
%%%%%%%%%%%%%%%%%%%%%%%%%%%

\section{$\mathcal{N}=1$ superconformal duality-invariant model} \label{Section2}

We consider a dynamical system describing an Abelian $\cN = 1$
vector multiplet in curved superspace and denote by $S[W , {\bar W}]$ 
the corresponding action functional. The action is assumed to depend
on the  chiral spinor field strength $W_\a$
 and its conjugate ${\bar W}_\ad$ which are constructed 
  in terms of a real unconstrained gauge prepotential $V$ \cite{FZ,WZ} as
\bea
W_\a = -\frac{1}{4}\,  (\bar \cD^2 -4R) 
 \cD_\a  V~, \qquad \bar \cD_\bd W_\a=0~.
\eea
The prepotential is defined modulo gauge transformations
\bea
\d_\l V = \l + \bar \l ~, \qquad \bar \cD_\ad \l =0~, 
\eea
such that $\d_\l W_\a =0$.
The gauge-invariant  field strengths $W_\a$ and ${\bar W}_\ad$  obey
the Bianchi identity
\bea
\cD^\a W_\a = \bar \cD_\ad {\bar W}^\ad~,
\label{eq:bianchi}
\eea
and thus $W_\a$ is a reduced chiral superfield. 
We assume that $S[W , {\bar W}]$ does not involve the combination $\cD^\a W_\a $ as an independent variable, and therefore it
can unambiguously be defined
as a functional of a {\it general} 
 chiral superfield $W_\a$ and its conjugate ${\bar W}_\ad$.
Then, defining 
\bea
{\rm i}\,M_\a := 2\, \frac{\d }{\d W^\a}\,S[W , {\bar W}]~,
\label{2.3M}
\eea
the equation of motion for $V $ is 
\bea
\cD^\a M_\a = \cDB_\ad {\bar M}^\ad~.
\label{eq:eom}
\eea
Here the variational derivative $\d S/\d W^\a $ is defined by 
\bea
\d S =  \int \rd^4 x \,{\rm d}^2 \q \,\cE\, \d W^\a \frac{\d S}{\d W^\a}~+~{\rm c.c.}~,
\eea
where $\cE$ denotes the chiral integration measure, and $W^\a$ is
assumed to be an unrestricted covariantly chiral spinor.

Since the Bianchi identity (\ref{eq:bianchi}) and the equation of
motion (\ref{eq:eom}) have the same functional form, one may
consider $\sU(1)$ duality rotations
\bea
\d W_\a = \l M_\a ~, 
\qquad \d M_\a = - \l W_\a~,
\label{DualRot}
\eea
with $\l \in {\mathbb R}$ a constant parameter. The condition for duality invariance is the so-called self-duality equation 
\bea
{\rm Im} \int \rd^4 x \rd^2 \q  \,\cE \Big\{ W^\a W_\a  +M^\a M_\a \Big\} =0~,
\label{SDE1}
\eea
in which $W_\a$ is chosen to be a general chiral spinor.

%%%%%%%%%%%%%%%%%%%%%%%%%%%%%%%
%%%%%%%%%%%%%%%%%%%%%%%%%%%%%%%

\subsection{Formulation without auxiliary chiral variables} 

General duality-invariant supersymmetric theories with at most two derivatives 
at the component level were constructed in \cite{KT1,KT2,KMcC,KMcC2}.
They belong to the family of  nonlinear vector multiplet theories of the general form 
\bea
S[W,{\bar W};\U] &=&
\frac{1}{4} \int  \rd^4 x \rd^2 \q  \,\cE \, W^2 +{\rm c.c.}
\non \\
&& + \frac14  \int \rd^4 x \rd^2 \q \rd^2\bar \q \,E \,
\frac{W^2\,{\bar W}^2}{\U^2}\,
\L\left(\frac{u}{\U^2},
\frac{\bar u}{\U^2}\right)~,
\label{2.1}
\eea
where $W^2 =W^\a W_\a$ and $\bar W^2 = \bar W_\ad \bar W^\ad$,  the complex variable $u$ is defined by
\bea
u  := \frac{1}{8} (\cD^2 - 4  \bar R)  W^2~,
\eea
and $\U$ is a nowhere vanishing real  scalar with the super-Weyl transformation 
\bea
\d_\s \U = (\s +\bar \s) \U~, \qquad \bar \cD_\bd \s=0~, 
\label{2.3}
\eea
with $\s$ being the super-Weyl parameter, see appendix \ref{AppendixB} for more details.
The transformation law \eqref{2.3} implies that \eqref{2.1} is super-Weyl invariant.
We remind  the reader that $V$ is inert under the super-Weyl
transformations,
\bea
\d_\s V =0 \quad \implies \quad \d_\s W_\a = \frac{3}{2} \s \, W_\a \quad \implies 
\quad \d_\s (\cD^\a W_\a ) = (\s +\bar \s) \cD^\a W_\a~,
\eea
and therefore the following composite chiral scalar 
\bea
(\cD^2 - 4 {\bar R}) \Big( \frac{W^2 }{ \U^2 } \Big)
\eea
is super-Weyl invariant. 
In the functional \eqref{2.1}, the expression in the first line is  the free vector multiplet action, while the interaction effects are encoded by 
 $\L(z,\bar z)$ which is a real function  of a complex variable $z$.

Three different realisations of $\U$ are possible:
\begin{enumerate} 
\item
One option is that 
$\U$  is a composite superfield, which is constructed in 
terms of the chiral compensator $S_0$ of old minimal supergravity and matter chiral superfields $\vf^i$,
\bea
\U    &=& S_0 \bar S_0 \,
{\rm exp} \Big(-\frac{1}{3} K (\vf^i,{\bar \vf}^{\bar j})\Big)~, \qquad \bar \cD_\bd S_0=0~,
\quad \bar \cD_\bd \vf^i =0~,
\eea
where $K(\vf,{\bar \vf})$ is the K\"ahler potential of a K\"ahler manifold.
The super-Weyl transformations laws of the chiral compensator and matter chiral superfields are 
\bea
\d_\s S_0 = \s S_0~, \qquad \d_\s \vf^i =0~.
\eea
\item
Another choice for $\U$ is the real linear compensator ${\mathbb L} $ of new minimal supergravity, 
\bea
(\bar \cD^2 - 4R) {\mathbb L} = (\cD^2 -4\bar R){\mathbb L} =0~.
\eea
This constraint is only compatible with the super-Weyl transformation law 
\bea
\d_\s {\mathbb L} = (\s + \bar \s) {\mathbb L} ~.
\eea
\item
One more option is given by $\U = \cD^\a W_\a = \bar \cD_\ad \bar W^\ad$, 
which corresponds to the family of superconformal vector multiplet models
introduced in \cite{K19}:
\bea
S [W, \bar W] &=&
\frac{1}{4}  \int \rd^4 x \rd^2 \q   \, \cE\,
W^2 +{\rm c.c.} \non \\
&&+ \frac 14 \int \rd^4 x \rd^2 \q  \rd^2 \bar{\q} \, E\,
\frac{W^2\,{\bar W}^2}{(\cD W)^2}\,
{\mathfrak H} \left( \frac{u  }{(\cD W)^{2}}
\,, \, \frac{ \bar u }{(\cD W)^{2}} \right)~.~~~~
\label{2.17}
\eea
Here ${\mathfrak H}(z,\bar z)$ is a real function of a complex variable.
In general,  this action explicitly depends on $\cD^\a W_\a $, and thus
there is no way to unambiguously define
it as a functional of an unrestricted  chiral spinor $W_\a$ 
and its conjugate ${\bar W}_\ad$; such models are not compatible with duality invariance. 
However, all dependence on  $\cD^\a W_\a $ disappears provided\footnote{The most general function ${\mathfrak H}(z,\bar z)$, for which $ \cD^\a W_\a $ drops from the action \eqref{2.17},  is given by  ${\bm {\mathfrak H}}_{\rm SC}(z, \bar z)
= (z \bar z)^{-1/2} f( u /\bar u) $, with $f$ a function on $S^1$. However, it is the special choice \eqref{2.18} which is compatible with the self-duality equation  \eqref{SDE1}.
}
\bea
{\mathfrak H}(z, \bar z) = \frac{y}{\sqrt{z\bar z} } +\hf (1-x) \Big( \frac{1}{z} +\frac{1}{\bar z} \Big)~,
\label{2.18}
\eea
for real parameters $x, y$.
\end{enumerate} 
In what follows, it will be assumed that $\U$ is a  compensating multiplet independent of the vector multiplet $V$, which excludes option 3 with the exception of \eqref{2.18}.

The model \eqref{2.1} is $\sU(1)$ duality-invariant if the interaction $\L(u,\bar u)$ 
satisfies the following differential equation 
\bea
{\rm Im} \,\Big\{ \G
- \bar{u}\, \G^2
\Big\} = 0~, \qquad 
\G  := \frac{\pa (u \, \L) }{\pa u}~, \qquad \L = \L(u,\bar u)~.
\label{differential}
\eea
The function $ \L(u,\bar u)$ was chosen in \cite{KT1,KT2} to be real analytic, in order for the model to be well-defined in the weak-field  limit. However $ \L(u,\bar u)$ is not required to be globally analytic.
It should be pointed out that setting  $\U=g^{-1} ={\rm const}$ in \eqref{2.1} and choosing
\bea
\L(u, \bar u) &=& \frac{1 }
{ 1 + \hf\, A \, +
\sqrt{1 + A +\frac{1}{4} \,B^2} }~,
\qquad  A =   u+\bar u~, \quad 
B = u-\bar u
\eea
defines the $\cN=1$ supersymmetric Born-Infeld action \cite{CF}.
This $\sU(1)$ duality-invariant theory is a Goldstone multiplet 
action for partial   $\cN=2 \to \cN=1$ supersymmetry 
breaking in Minkowski space \cite{BG,RT},  as well as in the following maximally supersymmetric backgrounds \cite{KT-M16}: 
(i)  ${\mathbb R} \times S^3$; (ii) ${\rm AdS}_3 \times {\mathbb R}$;
and (iii) a supersymmetric plane wave.  

%%%%%%%%%%%%%%%%%%%%%%%%%%%%%%
%%%%%%%%%%%%%%%%%%%%%%%%%%%%%%%

\subsection{Superconformal duality-invariant model}  \label{subsection2.2}

The model \eqref{2.1} is superconformal provided the functional form of $ \L(u,\bar u)$ is such that \eqref{2.1}  is independent of $\U$.\footnote{The vector multiplet action \eqref{2.17} is superconformal. However it is not compatible with duality invariance
since the integrand depends on $\cD^\a W_\a$.}
This condition implies that 
\bea
{\bm \L}_{\rm SC}(u, \bar u) = (u\bar u)^{-\hf} f(u/\bar u)~, 
\qquad f: S^1 \to {\mathbb R}~.
\label{221}
\eea
However, 
only a special choice of $f$  proves to be compatible with duality invariance, eq.  \eqref{differential}, 
specifically  
 \begin{subequations}\label{2.5}
\bea
\L_{\rm SC}(u, \bar u) = \frac{y}{\sqrt{u\bar u} } +\hf (1-x) \Big( \frac{1}{u} +\frac{1}{\bar u} \Big)~,
\eea
with $x$ and $y$ real parameters. 
Now the self-duality equation \eqref{differential}
is satisfied iff
\bea
x^2 - y^2 = 1 \quad \Longleftrightarrow \quad x =\cosh \g ~, \quad y = \sinh \g~.
\eea
\end{subequations}
Due to the identity 
\bea
 \int \rd^4 x \rd^2 \q \rd^2\bar \q \,E \,
\frac{W^2\,{\bar W}^2}{\bar u} = - 2  \int  \rd^4 x \rd^2 \q  \,\cE \, W^2 ~,
\eea
the superconformal $\sU(1)$ duality-invariant  action takes the form 
\bea
S[W,{\bar W}] &=&
\frac{1}{4} \cosh \g \int  \rd^4 x \rd^2 \q  \,\cE \, W^2 +{\rm c.c.}
\non \\
&& + \frac{1}{4}\sinh \g   \int \rd^4 x \rd^2 \q \rd^2\bar \q \,E \,
\frac{W^2\,{\bar W}^2}{\sqrt{u\bar u} }~.
\label{2.24}
\eea
This is the model proposed in \cite{BLST2}.

%%%%%%%%%%%%%%%%%%%%%%%%%%%%%%%
%%%%%%%%%%%%%%%%%%%%%%%%%%%%%%%%

\subsection{Formulation with auxiliary chiral variables}

There exists a different formulation for the models discussed above. 
Following \cite{K13}, we consider an action functional of the form 
\begin{subequations}\label{2.9}
\bea
S[W,\bar W, \eta, \bar \eta; \U]&=& \int{\rm d}^4 x \rd^2\q\,\cE \Big\{ \eta W -\hf \eta^2 - \frac{1}{4} W^2\Big\} 
+{\rm c.c.} \non \\
&& \quad+ \frac{1}{4} \int \rd^4 x \rd^2 \q \rd^2\bar \q \,E \, \frac{\eta^2 \bar \eta^2}{\U^2} 
{\mathfrak F}\Big( \frac{v}{\U^2} , \frac{\bar v}{\U^2} \Big)~,
\eea
in which
\bea
v:= \frac{1}{8} (\cD^2 -4\bar R) \eta^2~,
\eea
\end{subequations}
and the auxiliary spinor $\eta_\a$ 
is only constrained to be covariantly chiral, $\bar \cD_\bd \eta_\a =0$.
The action is super-Weyl invariant if $\eta_\a$ transforms as\footnote{This super-Weyl transformation law  
coincides with that of $W_\a$ and the chiral spinor prepotential of the tensor multiplet given in section 6.7 of  \cite{Ideas}.}
\bea
\d_\s \eta_\a = \frac{3}{2} \s \eta_\a~,
\eea
in conjunction with the transformation of $\U$, eq. \eqref{2.3}. 

Making use of the equations of motion for $\eta_\a$ and $\bar \eta_\ad$ allows one to integrate out these variable to end up with an action of the form \eqref{2.1}. 
For the interaction 
$\L(u, \bar u)$ one obtains 
\bea
\L(u,\bar u):= \frac{ \fF + \bar v [\pa_v(v\fF)]^2 + v[\pa_{\bar v}(\bar v \fF)]^2 }
{ \big[ 1+ \pa_v (v\bar v \fF)\big]^2 
\big[ 1+ \pa_{\bar v} (v\bar v \fF)\big]^2}~,
\eea
see \cite{K13} for the technical details. Here the variables $u$ and $v$ are related to each other as follows
\bea
u \approx v [1+\pa_v (v\bar v {\mathfrak F})]^2 ~,
\eea
where the symbol $\approx$ is used to indicate that the result holds modulo terms 
proportional to $\eta_\a $ and $\bar  \eta_\ad$ or, equivalently, 
to $W_\a $ and $\bar  W_\ad$ (such terms do not contribute to the action).

Our  model \eqref{2.9} possesses $\sU(1)$ duality invariance under the condition 
\bea
{\mathfrak F} ( v,  \bar v) 
={\mathfrak F} (v \bar v)~,
\label{2.26}
\eea
see \cite{K13} for the technical details. 
 
%%%%%%%%%%%%%%%%%%%%%%%%%%%%
%%%%%%%%%%%%%%%%%%%%%%%%%%%%

\subsection{Superconformal duality-invariant model} \label{subsection2.4}
 
The duality-invariant model defined by eqs. \eqref{2.9} and \eqref{2.26} 
is superconformal if the action is independent of $\U$, which means that the functional form of $\mathfrak F$ is fixed modulo a single real parameter, 
\bea
{\mathfrak F}_{\rm SC} ( v  \bar v) = \frac{\k}{\sqrt{v \bar v}}~.
\label{2.27}
\eea

The auxiliary variables $\eta_\a$ and $\bar \eta_\ad$ can be integrated out
using the equation of motion for  $\eta_\a$, 
\bea
W_\a = \eta_\a \left\{ 1 + \frac{1}{8} (\bar \cD^2 -4R) 
\Big[ {\bar \eta}^2 \Big({\mathfrak F}_{\rm SC}  + \frac{1}{8} (\cD^2 -4\bar R) 
\big( \eta^2\, \pa_v {\mathfrak F}_{\rm SC} \big) \Big) \Big] \right\}~,
\eea
and its conjugate.
Rather lengthy calculations lead to
\bea
S[W,{\bar W}] &=&
\frac{1}{4} \int  \rd^4 x \rd^2 \q  \,\cE \, W^2 +{\rm c.c.}
\non \\
&& + \frac14  \int \rd^4 x \rd^2 \q \rd^2\bar \q \,E \,
W^2\,{\bar W}^2\,
\L_{\rm SC} (u,\bar u)~,
\eea
where $\L_{\rm SC} (u,\bar u)$ has the form \eqref{2.5}, with the parameters $x$ and $y$ being given by 
\bea
x=\frac{1+(\k/2)^2}{1-(\k/2)^2}~, \qquad y=\frac{\k}{1-(\k/2)^2}~.
\label{2.31}
\eea 

It is interesting to compare the results obtained in subsections \ref{subsection2.2}
and  \ref{subsection2.4}. Within the approach without auxiliary chiral variables, 
the most general 
superconformal coupling is given by eq. \eqref{221} and involves an arbitrary real function 
$f$ on $S^1$. Requiring the superconformal coupling to obey the self-duality equation
\eqref{differential} fixes the functional form of $\L_{\rm SC}(u, \bar u) $ to be given by eq. \eqref{2.5}. The remaining freedom in the choice of $\L_{\rm SC}(u, \bar u)$ is the single 
real parameter $\k$. On the other hand, when the formulation with auxiliary chiral variables is used, duality invariance is guaranteed by the condition \eqref{2.26}. Upon imposing this condition, there exists a unique functional form for superconformal coupling,  eq. \eqref{2.27}, and the only remaining freedom is again the coupling constant $\k$.

%%%%%%%%%%%%%%%%%%%%%%%%%%
%%%%%%%%%%%%%%%%%%%%%%%%%%%

\subsection{Superconformal invariance} 

To conclude this section, we comment on rigid symmetries of the superconformal models \eqref{2.17}, including the duality-invariant theory \eqref{2.24}, in a background curved superspace. We remind the reader that, within the superconformal approach to supergravity-matter systems \cite{KakuT}, 
every theory of Einstein supergravity interacting with supersymmetric matter is realised as a coupling of the same matter multiplets to conformal supergravity and a superconformal compensator, see, e.g., 
\cite{FGKV}. Truly superconformal theories are independent of any compensator.

The gauge group of conformal supergravity is spanned by general coordinate  ($\xi^{B}$), local Lorentz ($K^{ \b \g}$ and $\bar K^{\bd \gd}$)
 and super-Weyl ($\s$ and $\bar \s$) transformations. In the infinitesimal case, they act on the covariant derivatives by the rule
\bea
\d \cD_A = \delta_{\mathcal{K}} \cD_{A} +\d_\s \cD_A ~,
\label{2.35}
\eea
where the general coordinate and local Lorentz transformation is given by 
\bea
\delta_{\mathcal{K}} \cD_{A} = \left[ \mathcal{K} , \cD_{A} \right] ~, 
\qquad \mathcal{K} = \xi^{B} \cD_{B} + K^{\b\g} M_{ \b \g} 
+ \bar K^{\bd \gd} \bar M_{\bd \gd}  =\bar \cK
\eea
and the super-Weyl variation $\d_\s \cD_A$ is defined in \eqref{superweyl}.
The local transformation \eqref{2.35} acts on a primary tensor superfield 
${\mathfrak T}$ (with its  indices suppressed) as follows 
\bea
\d {\mathfrak T} = \d_\cK  {\mathfrak T} + \d_\s  {\mathfrak T} 
=\cK {\mathfrak T} +(p\s +q \bar \s) {\mathfrak T}~.
\label{2.37}
\eea
Here  $p$ and $q$ are constant parameters which are known as the super-Weyl weights of $ {\mathfrak T}$, with $(p+q)$ being the dimension of 
$ {\mathfrak T} $.
The action of any supergravity-matter system is invariant under the transformations \eqref{2.35} and \eqref{2.37}, where ${\mathfrak T} $ is the collective notation for the matter multiplets and, perhaps, the compensator (the latter is present if the theory is not superconformal).

Let us consider a background curved superspace $(\cM^{4|4}, \cD) $. 
In accordance with section 6.4 of\cite{Ideas}, 
a supervector field $\x= \x^B E_B$ on $(\cM^{4|4}, \cD)$
is called conformal Killing if there exists a 
symmetric spinor $K^{\g\d}$ and a covariantly chiral scalar $\s$ such that 
\bea
(\d_\cK + \d_\s) \cD_A =0~.
\label{conf-Killing}
\eea
In other words, the  coordinate transformation generated by $\x$ can be accompanied by certain Lorentz and super-Weyl transformations such that the superspace geometry does not change. It turns out that the parameters $\x^\a$, $K^{\a\b}$ and $\s$
are completely determined in terms of  $\xi^{a}$ and its covariant derivatives, 
\begin{subequations} \label{2.39ab}
\bea
 { \x}^\a &=& -\frac{ \ri }{8} \bar \cD_\ad \x^{\a \ad}
~, \qquad 
K^{\a\b} [\x]= \cD^{(\a } \x^{\b)}  +\frac{\ri}{2} \x^{ (\a}{}_\bd G^{\b) \bd}~, \\
\s[\x] &=&- \frac{1}{3} (\cD^\a \x_\a +2 \bar \cD_\ad \bar \x^\ad - {\ri}  \x^{a} G_{a} )~,
\eea
\end{subequations}
The real vector $\x^a$ proves to obey the superconformal Killing equation
\bea
\cD_{(\a} \x_{\b) \bd} =0 \quad \Longleftrightarrow \quad \bar \cD_{(\ad} \x_{\b \bd)} =0~.
\label{master}
\eea
We denote by $\d_{\cK [\x]} $ and $\d_{\s [\x]}$ the transformations
associated with the conformal Killing supervector field $\xi^{A} = \big( \xi^{a} , - \frac{\rm i}{8} \cDB_{\bd} \xi^{\a \bd} , 
- \frac{\rm i}{8} \cD^{\b} \xi_{\b \ad} \big)$. The set of all conformal Killing supervector fields forms the superconformal algebra of $(\cM^{4|4}, \cD) $. 
If the superspace is conformally flat (for instance, Minkowski ${\mathbb M}^{4|4}$ or anti-de Sitter  AdS$^{4|4}$ superspace), its superconformal algebra is isomorphic to
$\mathfrak{su}(2,2|1)$. 

Given a superconformal field theory defined on the background  superspace $(\cM^{4|4}, \cD) $, its action functional  is invariant under superconformal transformations 
\bea
\d_\x {\mathfrak T}  = (\d_{\cK [\x]} + \d_{\s [\x]}){\mathfrak T} ~.
\eea
However, if the theory under consideration is not superconformal, then the set 
${\mathfrak T}  $ also includes a conformal compensator $\X$ which is non-dynamical  and, instead, it is a part of the background supergravity multiplet  $(\cM^{4|4}, \cD, \X) $. 
In this case the equations \eqref{2.39ab} and \eqref{master} must be accompanied by 
the additional condition
\bea
(\d_{\cK [\x]} + \d_{\s [\x]}) \X=0~,
\eea
which singles out the Killing supervector fields of  $(\cM^{4|4}, \cD, \X) $. 

The above discussion can be naturally modified to be applicable to the $\cN=2$ superconformal theories studied in the next section. Similar symmetry considerations
also hold for supergravity-matter  theories in $d\leq 6$ 
with up to eight supercharges, where off-shell conformal supergravity always exists, see
\cite{K15Corfu} for the technical details.  

%%%%%%%%%%%%%%%%%%%%%%%%%%%%
%%%%%%%%%%%%%%%%%%%%%%%%%%%%

\section{$\mathcal{N}=2$ superconformal duality-invariant models}\label{Section3}

Let $ S[W , {\bar W}]$ be the action for an Abelian $\cN=2$ vector multiplet 
coupled to  supergravity. As a curved-superspace extension of the Grimm-Sohnius-Wess formulation \cite{GSW}, 
the vector multiplet is described by a covariantly chiral scalar superfield $W$, 
\bea
\cDB_{\ad i} W= 0~, 
\label{3}
\eea
  subject to the Bianchi identity
\bea
\Big( \cD^{ij} + 4S^{ij}\Big) W&=&
\Big(\bar \cD^{ij} +  4\bar{S}^{ij}\Big)\bar{W} ~,
\label{n=2bi-i}
\eea
where  $\cD^{ij}:= \cD^{\a(i}\cD_\a^{j)}$ and 
$\bar \cD^{ij} :=  \cDB_\ad{}^{(i}\cDB^{j) \ad}$, while 
$S^{ij} $ and its conjugate ${\bar S}^{ij}$  
are special  components of the superspace torsion.
Constraint \eqref{n=2bi-i} defines a reduced $\cN=2$ chiral scalar superfield. 
The superspace formulation for $\cN=2$ 
conformal supergravity developed in \cite{KLRT-M} is used in this section, see appendix 
\ref{AppendixC} for the technical details.

The reduced chiral scalar $W$ is a gauge-invariant field strength
constructed in terms of Mezincescu's prepotential \cite{Mezincescu,HST}, 
  $V_{ij}=V_{ji}$,
which is an unconstrained real SU(2) triplet, 
$\overline{V_{ij}} = V^{ij}= \ve^{ik}\ve^{jl}V_{kl}$. 
The expression for $W$ in terms of $V_{ij}$ 
was found in \cite{ButterK} to be
\begin{align}
W = 
\bar\Delta \Big({\cD}^{ij} + 4 S^{ij}\Big) V_{ij}~.
\label{Mez}
\end{align}
Here $\bar\Delta$ is the covariantly chiral projecting operator \eqref{chiral-pr}.
The prepotential  $V_{ij}$ is defined modulo gauge transformations \cite{ButterK2}
\bea
\delta_\L V^{ij} &= \cD^{\alpha}_k \Lambda_\alpha{}^{kij}
     + \bar\cD_{\dalpha}{}_k \bar\Lambda^\dalpha{}^{kij}, \qquad
     \Lambda_\alpha{}^{kij} = \Lambda_\alpha{}^{(kij)}~,
     \qquad \bar\Lambda^\dalpha{}_{kij} := 
     \overline{ \Lambda_\alpha{}^{kij} }~,
\label{pre-gauge1}
\eea
with the gauge parameter $ \Lambda_\alpha{}^{kij} $ being arbitrary modulo 
the algebraic condition given.\footnote{In the flat-superspace limit, the gauge transformation law \eqref{pre-gauge1} reduces to that given in 
\cite{Mezincescu}.}
It is an instructive exercise to show that $\d_\L W=0$.

The super-Weyl transformation law of $V_{ij}$   \cite{K12} is
\bea
\d_\s V_{ij} = -(\s +\bar \s) V_{ij}~.
\eea 
Making use of the relations collected in appendix \ref{AppendixC}, 
we then deduce the super-Weyl transformation law of the field strength $W$ 
\cite{KLRT-M}
\bea
\d_{\s} W = \s W~.
\label{Wsuper-Weyl}
\eea

%%%%%%%%%%%%%%%%%%%%%%%%%%%%
%%%%%%%%%%%%%%%%%%%%%%%%%%%%%

\subsection{Formulation without auxiliary superfields} 

To describe the equations of motion, 
we introduce a covariantly chiral scalar superfield
 $M$ defined as\footnote{It is assumed here that $S[W , {\bar W}]$ is consistently defined as a functional of a general $\cN=2$ chiral scalar $W$ and its conjugate.}
\be
{\rm i}\, M := 4\, \frac{\d }{\d W}\,
S[W , {\bar W}]
~,  \qquad \cDB_{\ad i} M= 0~. 
\label{n=2vd}
\ee
In terms of $M$ and its conjugate $\bar M$,  the equation of motion for $V_{ij}$ is
\bea
\Big(\cD^{ij}+4S^{ij}\Big) M&=&
\Big(\cDB^{ij}+4\bar{S}^{ij}\Big)\bar{M} ~.
\label{n=2em}
\eea
Making use of \eqref{Wsuper-Weyl} and 
the super-Weyl transformation of the chiral density \cite{KLRT-M}, eq. \eqref{C.8},
we obtain the super-Weyl transformation of $M$:
\bea
\d_{\s} M = \s M~.
\eea
We see that $W$ and $M$ have the same super-Weyl transformation.

Since the Bianchi identity (\ref{n=2bi-i}) and the equation of
motion (\ref{n=2em}) have the same functional form,
one can consider infinitesimal $\sU(1)$ duality rotations
\be
\d W = \l \, M~, \qquad 
\d M  = -\l \, W~,
\label{n=2dt}
\ee
with $\l \in \mathbb R$ a constant parameter. The theory under consideration is 
duality invariant under the condition  \cite{K12} 
\bea
{\rm Im} \int \rd^4 x \,{\rm d}^4\q \,\cE\,  \Big( W^2 + M^2 \Big) =0~.
\label{SDE}
 \eea
In the rigid superspace limit, this reduces to the $\cN=2$ self-duality equation
derived in \cite{KT1}.

%%%%%%%%%%%%%%%%%%%%%%%%%%%%
%%%%%%%%%%%%%%%%%%%%%%%%%%%%%%%

\subsection{Formulation with auxiliary chiral variables}

The duality-invariant models discussed in the previous subsection possess an important reformulation  \cite{K13}.
We consider a locally supersymmetric theory with action
\bea
S[W,\bar W, \eta, \bar \eta]= \hf \int \rd^4 x \rd^4 \q  \,\cE \Big\{ \eta W -\hf \eta^2 - \frac{1}{4} W^2\Big\} 
+{\rm c.c.}  + {\mathfrak S}_{\rm int} [\eta, \bar \eta]~,
\label{3.8}
\eea
in which the scalar superfield $\eta$ is only constrained to be chiral, $\bar \cD_\bd \eta =0$. 
We require $S[W,\bar W, \eta, \bar \eta]$ to be super-Weyl invariant, and therefore 
the corresponding transformation of $\eta $ is\footnote{This super-Weyl transformation law  
coincides with that of $W$ and the chiral scalar prepotential of the $\cN=2$ tensor multiplet \cite{Kuzenko:2008qz}.}  
\bea
\d_{\s} \eta = \s \eta~.
\eea
For ${\mathfrak S}_{\rm int} [\eta, \bar \eta]$ to be super-Weyl invariant, 
\bea
\d_\s {\mathfrak S}_{\rm int} [\eta, \bar \eta]=0~,
\eea
it may  depend, in general, on the supergravity compensators. However, in this paper we are mostly interested in superconformal dynamical systems
which possess no dependence on the compensators.

We assume that $\eta$ and its conjugate $\bar \eta$ are auxiliary  
superfields in the sense that the equation of motion for $\eta$,
\bea
W= \eta- 2\frac{\d}{\d \eta}  {\mathfrak S}_{\rm int} [\eta, \bar \eta]~,
\eea
and its conjugate may be solved, at least in perturbation theory, to express $\eta$ as a functional 
of the field strength $W$ and its conjugate, $\eta =\eta [W, \bar W]$.
As a result, we end up with the action 
\bea
S[W,\bar W] = S[W,\bar W, \eta, \bar \eta]\Big|_{\eta =\eta [W, \bar W]}~,
\eea
which describes the dynamics of the vector multiplet. 

The above action, eq. \eqref{3.8},  defines a $\sU(1)$ duality-invariant system provided  
${\mathfrak S}_{\rm int} [\eta, \bar \eta]$ is invariant under $\sU(1)$ rigid transformations,
\bea
{\mathfrak S}_{\rm int} [ \re^{\ri \vf}  \eta, \re^{-\ri \vf} \bar \eta] 
={\mathfrak S}_{\rm int} [\eta, \bar \eta] ~, 
\qquad \vf \in {\mathbb R}~.
\label{3.4}
\eea

%%%%%%%%%%%%%%%%%%%%%%%%%%%%%
%%%%%%%%%%%%%%%%%%%%%%%%%%

\subsection{Superconformal duality-invariant models}

We introduce the super-Weyl invariant functional
\bea
2 {\mathfrak S}_{\rm int} [\eta, \bar \eta] &=&\int \rd^4x\, \rd^4\q\, \cE\,  \Big\{  (c-a)
W^{\a\b}W_{\a\b} 
+ a  \X \Big\}\ln \eta   ~+~{\rm c.c.} \non \\
&&  +2 a \int \rd^4 x \,{\rm d}^4\q\,{\rm d}^4{\bar \q}\,E\, \ln \eta \ln \bar \eta
-\G_{\rm eff} ~,
\label{3.5}
\eea
where $W_{\a\b}$ is the super-Weyl tensor, the chiral scalar $\X$ 
is given by eq. \eqref{C.7}, and 
$a$ and $c$ are real anomaly coefficients.
In accordance with \cite{K13anomaly},
$\G_{\rm eff}$ is a nonlocal functional of the background supergravity multiplet
which generates the super-Weyl anomaly, with its super-Weyl variation being 
\bea
\d_\s \G_{\rm eff} 
 = \int \rd^4x\, \rd^4\q\, \cE\, \s \Big\{  (c-a)W^{\a\b}W_{\a\b} 
+a \X \Big\}~+~{\rm c.c.} ~~
\eea
By construction, $\G_{\rm eff}$ is independent of $\eta $ and $\bar \eta$.

The functional \eqref{3.5} satisfies the  condition \eqref{3.4} 
due to the following two properties. Firstly, for any covariantly chiral scalar $\f$,
it holds that 
 \bea
{\bar \cD}^{\ad}_i  \f=0
\quad \Longrightarrow \quad 
\int \rd^4 x \,{\rm d}^4\q\,{\rm d}^4{\bar \q}\,E\, \f=0~.
\eea
Secondly, the functionals
\begin{subequations}
\bea
{\mathfrak T}_1&=& \int \rd^4x\, \rd^4\q\, \cE\, \Big\{ W^{\a\b}W_{\a\b}
-\X\Big\} ~, \label{3.20a} \\
{\mathfrak T}_2&=&{\rm Im}  \int \rd^4x\, \rd^4\q\, \cE\,  W^{\a\b}W_{\a\b} 
\eea
\end{subequations}
are topological invariants. In the case of $\cN=4$ SYM theory, the anomaly coefficients 
$a$ and $c$ coincide, $a=c$. In what follows, we will assume this relation and make use of $c$.

Another family of $\sU(1)$-invariant couplings is given by 
\bea
\widetilde {\mathfrak S}_{\rm int} [\eta, \bar \eta] = \int \rd^4 x \,{\rm d}^4\q\,{\rm d}^4{\bar \q}\,E\,  
{\mathfrak F} ({ \O} \bar { \O}) ~, 
\eea
where we have introduced the super-Weyl inert chiral superfield
\bea
{ \O} := \frac{ 1}  {\eta^2} \Big(  \bar \D \ln \bar \eta + \hf  \X \Big) ~, \qquad 
\bar \cD^\ad_i { \O} =0~, \qquad \d_\s { \O} =0~.
\label{Omega}
\eea
This is a curved superspace generalisation of the conformal primary weight-zero 
chiral superfield introduced in \cite{BKT}. 
The superfield $\eta^2 { \O}$ naturally 
originates within the approach developed in \cite{BdeWKL}. 

With the contribution $(-\G_{\rm eff} )$ included, the action defined by eqs. \eqref{3.8}
and \eqref{3.5} is super-Weyl invariant but nonlocal. Instead of working with the nonlocal functional, we can consider the following $\sU(1)$ duality-invariant action 
\bea
S
&=& {\rm Re} \int \rd^4 x \rd^4 \q  \,\cE \Big\{ \eta W -\hf \eta^2 - \frac{1}{4} W^2 +
\Big((c-a) W^{\a\b}W_{\a\b} +a \X \Big) \ln \eta\Big\} 
\non \\
&&+
a\int \rd^4 x \,{\rm d}^4\q\,{\rm d}^4{\bar \q}\,E\, \ln \eta \ln \bar \eta~.
\label{candidate}
\eea
This action generates the $\cN=2$ super-Weyl anomaly, 
and therefore it can be viewed as the Goldstone multiplet action for spontaneously broken $\cN=2$ superconformal symmetry.\footnote{The  $\cN=1$ dilaton effective action was proposed in \cite{SchT}, see also \cite{KST}.} 
It was proposed in \cite{K13anomaly} to describe
the $\cN=2$ dilaton effective action in terms of the reduced chiral scalar $W$. 
An alternative realisation was put forward in Ref. \cite{GHKSST} where 
the $\cN=2$ Goldstone multiplet  was identified with an unrestricted chiral scalar, such as $\eta$. 
Both superfields are present in \eqref{candidate}.

%%%%%%%%%%%%%%%%%%%%%%%%%%%%%%
%%%%%%%%%%%%%%%%%%%%%%%%%%%%%%

\subsection{Superconformal duality-invariant models in flat space}

In the flat-superspace limit, 
the supergravity covariant derivatives $\cD_A$, eq. \eqref{C.1}, turn into the flat ones
$D_A = (\pa_a, D^i_\a,
{\bar D}^\ad_i )$, with $i = {\1}, {\2}$, which satisfy the graded commutation relations
\bea
\{  D^i_\a ,D^j_\b  \} =
\{  {\bar D}_{\ad i},
{\bar D}_{\bd j}  \} =0~, \qquad
\{ D^i_\a , {\bar D}_{\ad j} \} =
-2{\rm i} \d^i_j (\s^a)_{\a \ad} \pa_a ~.
\eea

In Minkowski superspace,  the duality-invariant model \eqref{candidate} for the  $a=c$ choice, which corresponds to the $\cN=4$ SYM,
takes the form 
\bea
S[W,\bar W, \eta, \bar \eta]&=& 
{\rm Re} 
\int \rd^4 x \rd^4 \q  \, \Big\{ \eta W -\hf \eta^2 - \frac{1}{4} W^2\Big\} \non \\
&&+c\int \rd^4 x \,{\rm d}^4\q\,{\rm d}^4{\bar \q}\, \ln \eta \ln \bar \eta~.
\label{3.18} 
\eea
This action is invariant under standard $\cN=2$ superconformal transformations, see \cite{BKT} for the technical details. 
The equation of motion for $\eta$ is 
\bea
\eta = W +  \frac{2c}{\eta} \bar \D \ln \bar \eta ~.
\label{3.19}
\eea
Assuming the validity of \eqref{3.19}, the action \eqref{3.18} turns into 
\bea
S[W,\bar W]&=& \frac 14 {\rm Re} \int \rd^4 x \rd^4 \q  \, W^2
+c\int \rd^4 x \,{\rm d}^4\q\,{\rm d}^4{\bar \q}\, \ln \eta \ln \bar \eta \non \\
&&- c^2\int \rd^4 x \,{\rm d}^4\q\,{\rm d}^4{\bar \q}\, \left\{ 
\frac{1}{\eta^2} \ln \bar \eta \bar \D \ln \bar \eta + \frac{1}{\bar \eta^2} \ln  \eta  \D \ln  \eta
\right\}~,
\label{3.20}
\eea
where $\eta$ is assumed to be a functional of $W$ and its conjugate, 
$\eta = \eta[W, \bar W]$, obtained by solving the equation \eqref{3.19}. The latter can be solved by iterations as a series in powers of $c$, 
\bea
\eta = W +\frac{2c}{W} \bar \D \ln \bar W + \cO(c^2)~.
\eea
Making use of this solution in \eqref{3.20} gives
\bea
S[W,\bar W]&=& \frac 14 {\rm Re} \int \rd^4 x \rd^4 \q  \, W^2
+c\int \rd^4 x \,{\rm d}^4\q\,{\rm d}^4{\bar \q}\, \ln W \ln \bar W \non \\
&&+ c^2\int \rd^4 x \,{\rm d}^4\q\,{\rm d}^4{\bar \q}\, \left\{ 
\frac{1}{W^2} \ln \bar W \bar \D \ln \bar W + \frac{1}{\bar W^2} \ln W \D \ln W
\right\} +\cO(c^3)~.~~~~
\label{3.22}
\eea
This agrees with the perturbative solution derived in \cite{KT1}.

It is interesting to point out that the equation \eqref{3.19} is analogous to the 
$\cN=1$ chiral constraint (in the notation of \cite{KT2}) 
\bea
X +\frac 14 X {\bar D}^2 {\bar X} = W^\a W_\a~,
\eea
which is at the heart of the Bagger-Galperin construction \cite{BG} of the vector Goldstone multiplet action for partial   $\cN=2 \to \cN=1$ supersymmetry breaking in Minkowski space.  

The kinetic terms in \eqref{3.22} is purely classical. As  is known, it receives no perturbative \cite{deWGR,DeGiovanni:1997ib,BKO} and non-perturbative  
\cite{DKMSW, Matone}
quantum corrections. 
The $c$-term in \eqref{3.22}, originally introduced in \cite{deWGR}, contains
four-derivative quantum corrections at the component level; 
these include  an $F^4$ term, with $F$ the $\sU(1)$ field strength.
This term is  believed to be generated only 
at one loop in the $\cN=4$ SYM theory \cite{DS}. 
In particular it is known that non-perturbative 
$F^4$ quantum corrections
do not occur in this theory \cite{DKMSW,Matone}. 
The one-loop  calculation of $c$  was carried out by several groups
\cite{GKPR, PvU, G-RR, BK98, BBK,LvU}, and the absence of two-loop corrections 
was shown in \cite{BKO}.

The $c^2$-term in \eqref{3.22}, which was originally introduced in \cite{GKPR}
and later proved to be superconformal \cite{BKT},  contains
six-derivative quantum corrections at the component level,  including  an $F^6$ term.
Such a quantum correction is absent at one loop in the $\cN=4$ SYM theory \cite{BKT}.
It is generated at the two-loop order and the numerical coefficient computed in 
 \cite{K2004} agrees with that predicted by duality invariance.\footnote{A different value for the $F^6$-term coefficient  was derived in \cite{BPT}. Unfortunately,  an error was made in 
 \cite{BPT} in the process of evaluating the harmonic supergraphs, as explained in 
 \cite{K2004}.}

Our superconformal duality-invariant model \eqref{3.18} admits a natural generalisation 
described by the following action
\bea
S[W,\bar W, \eta, \bar \eta]&=&
{\rm Re}
\int \rd^4 x \rd^4 \q  \, \Big\{ \eta W -\hf \eta^2 - \frac{1}{4} W^2\Big\} 
\non \\
&&+\int \rd^4 x \,{\rm d}^4\q\,{\rm d}^4{\bar \q}\, \Big\{ c\ln \eta \ln \bar \eta
+ {\mathfrak F} ( \O \bar \O ) \Big\}~, 
\label{3.23} 
\eea
where ${ \O} := \eta^{-2}   \bar \D \ln \bar \eta $ is the flat-superspace 
counterpart of \eqref{Omega}. Here the functional in the second line is invariant under 
rigid $\sU(1)$ transformations
\bea
 \eta \to \re^{\ri \vf}  \eta~ , \qquad \bar \eta \to \re^{-\ri \vf} \bar \eta~,
\qquad \vf \in {\mathbb R}~.
\eea
The action \eqref{3.23} includes higher-derivative structures as compared with 
\eqref{3.18}. 

Modulo higher derivative terms, we identify  \eqref{3.18} with (the planar part of) the low-energy effective action of $\cN=4$ SYM on the Coulomb branch. 
There are several reasons for this identification. Firstly, upon elimination of the auxiliary variables $\eta $ and $\bar \eta$, the resulting action \eqref{3.22} agrees with known structure of the $\cN=4$ SYM effective action to the sixth order in derivatives. 
Secondly, the theory \eqref{3.18} possesses $\sU(1)$ duality invariance. 
Thirdly, the action \eqref{3.18} involves a single coupling constant, the anomaly coefficient $c$, which is similar to the D3-brane action in $AdS_5 \times S^5$, eq. 
\eqref{4.1}. The latter is expected to be related to he low-energy effective action of $\cN=4$ SYM (see the discussion in the next section), however only upon a nontrivial
field redefinition at the component level \cite{GKPR,K2004}. It remain possible that one has to include a $c$-dependent higher-derivative term ${\mathfrak F} ( \O \bar \O ) $ in \eqref{3.23} in order to get a complete agreement between the model. This is still an open 
problem.

%%%%%%%%%%%%%%%%%%%%%%%%%%%%
%%%%%%%%%%%%%%%%%%%%%%%%%%%%%

\section{Discussion}\label{Section4}

Similar to its bosonic sector \eqref{A.6}, 
the $\cN=1$ superconformal duality-invariant model is uniquely defined by eq. \eqref{2.24}   or, equivalently, eqs. 
\eqref{2.9} and \eqref{2.27}. The only free parameter of the theory is the coupling constant $\k$ in \eqref{2.27}. In the $\cN=2$ case, however, the superconformal duality-invariant 
model is given by the action \eqref{3.23}, which involves an arbitrary function of a  real variable, ${\mathfrak F}(x)$. The reason why the functional freedom is larger in the $\cN=2$ case can be explained by looking at the $\cN=1$ components of the $\cN=2$
vector multiplet field strength $W$.
Given an $\cN=2$ superfield $U(x, \q_i, \bar \q^i) $, its $\cN=1$ sub-multiplets can be 
introduced with the aid of 
the $\cN=1$ bar-projection defined by $ U| = U(\q_i , \bar \q^i)|_{ \q_{ \underline{2} } =
{\bar \q}^{{\2}} = 0}$. The $\cN=2$ chiral field strength $W$ 
contains two independent chiral $\cN=1$ components
\bea
\sqrt{2} \F := W | ~, \qquad 2{\rm i} W_\a:= D_\a^{\underline{2}}\, W |\quad \implies \quad (D^{\underline{2}})^2W|
=\sqrt{2} \, \bar{D}^2\bar\F~.
\eea
Modulo certain technical complications related to the need to perform a nonlinear superfield redefinition \cite{GKPR, K2004} (when switching from the $\cN=2$ to $\cN=1$ superfield descriptions),
the  $\cN=1$ counterpart of the system \eqref{3.23}
is given by the relations \eqref{2.9} and \eqref{2.26}, with $\U \propto \F \bar \F$, modulo terms involving derivatives of $\F$ and $\bar \F$.
The model for $\cN=1$ superconformal duality-invariant electrodynamics \eqref{2.24}  does not appear to have a $\cN=2$ counterpart. 

The AdS/CFT correspondence provides 
the main evidence to believe in self-duality of 
the low-energy effective action for the $\mathcal{N}=4$ $\mathsf{SU}(N)$ SYM theory on its Coulomb branch where the gauge group $\mathsf{SU}(N)$ is spontaneously broken to $\mathsf{SU}(N-1) \times \mathsf{U}(1)$.
The AdS/CFT correspondence  predicts 
\cite{Maldacena,CT,Tseytlin,BPT}
(a more comprehensive  list  of references is given  in \cite{BPT}) that the $\cN=4$ 
SYM effective action (in the large-$N$ limit)  is related to the 
D3-brane action in $AdS_5 \times S^5$
\bea
S&=& T_3 \int {\rm d}^4x \left( 
h^{-1} - \sqrt{ -\det ( g_{mn}  
+ F_{mn} ) } \right)~, \non \\
g_{mn} &=& h^{-1/2} \eta_{mn}
+h^{1/2}\,  \pa_m X^I \pa_n X^I ~,
\qquad h = {Q  \over (X^I X^I)^2 }~,
\label{4.1}
\eea
where $X^I$, $I=1,\cdots, 6$, are transverse 
coordinates,  
$T_3 =(2\p g_s)^{-1}$ 
and  $Q= g_s (N-1)/ \p$.
The action $S/T_3$ is self-dual in the sense 
that it enjoys invariance under 
electromagnetic $\sU(1)$ duality rotations  \cite{GR1,GR2,GZ2,GZ3}. 
Self-duality of the D3-brane action 
is a fundamental property 
related to the S-duality of type IIB string theory
\cite{Tseytlin96,GG}.
If the $\cN=4$ SYM effective action and 
the D3-brane action in $AdS_5 \times S^5$
are indeed related, the former should possess some form of self-duality.
The $\sU(1)$ duality-invariant action \eqref{4.1} involves a single coupling constant, $Q$, the same as the $\cN=2$ superconformal $\sU(1)$ duality-invariant action 
\eqref{3.18}. The D3-brane parameter $Q$ is proportional to the anomaly coefficient $c$.

Considered as a field theory in Minkowski space ${\mathbb M}^4$, the D3-brane action 
\eqref{4.1} is manifestly scale-invariant, but it is not invariant under the standard linear special conformal transformations. It is instead invariant under deformed conformal transformations being nonlinear in the fields \cite{Maldacena}.
At the component level, the complete effective action of the $\cN=4$ SYM also proves to be invariant under quantum-corrected 
superconformal transformations \cite{JKY,KM,KMT} which differ
from the ordinary linear superconformal  transformations
that leave  the classical action invariant.

In conclusion, we describe simple generalisations of the models for (super)conformal duality-invariant electrodynamics discussed in this paper.
Given a $\sU(1)$ duality-invariant model for nonlinear electrodynamics, $L(F_{ab}) =L(\o, \bar \o) $,
described in appendix \ref{AppendixA}, 
its  compact duality group $\sU(1)$
can be enhanced to the non-compact $\sSL(2,{\mathbb R})$ group by coupling the electromagnetic field to the dilaton $\vf$ and axion $\mathfrak a$ 
\cite{GR2,GZ2,GZ3}. Specifically, this is achieved by replacing $L(\o, \bar \o)  $, eq. \eqref{A.2}, with\footnote{The Lagrangian \eqref{43} can also be rewritten in the following form: 
$L(\o, \bar \o , \cS , \bar \cS) = -\frac{\ri}{2} (\bar \cS\o - \cS \bar \o) 
-\frac 14 (\bar \cS- \cS)^2 \o \bar \o \L \Big(  \frac{\ri}{2} (\bar \cS - \cS) \o , 
 \frac{\ri}{2} (\bar \cS - \cS) \bar \o \Big)$, compare with \eqref{4.10}.
}
\bea
L(\o, \bar \o , \cS , \bar \cS) = L(\cS_2 \o, \cS_2 \bar \o)  
+\frac{\ri}{2} \cS_1 ( \bar \o -\o) ~, \qquad 
\cS = \cS_1 + {\rm i}\, \cS_2 = {\mathfrak a} + {\rm i} \,{\rm e}^{-\vf}~.
\label{43}
\eea
The duality group acts by the rule
\bea
  \left( \begin{array}{c} G'  \\  F'  \end{array} \right)
=  \left( \begin{array}{cc} a~& ~b \\ c~ & ~d \end{array} \right) \;
\left( \begin{array}{c} G \\ F  \end{array} \right) ~, \qquad
\cS' = \frac{a\cS +b}{c\cS+d}~, \qquad
\left( \begin{array}{cc} a~& ~b \\ c~ & ~d \end{array} \right) \in
\sSL(2, {\Bbb R})~,
\label{A.10}
\eea
where 
$G_{ab}(F)$ is defined by 
\bea
\tilde{G}_{ab} (F)\equiv
\hf \, \ve_{abcd}\, G^{cd}(F) =
2 \, \frac{\pa L(F)}{\pa F^{ab}}~,
\eea
see \cite{KT2} for more details. 
In the case of the conformal model described by eq. \eqref{A.6}, 
this leads to 
\bea
L_{\rm conf}(\o, \bar \o , \cS , \bar \cS) =
- \hf \cosh \g  \Big( {\o} + {\bar \o}\Big) \cS_2
+ {\sinh \g }{\sqrt{\o\bar \o} }\cS_2 +\frac{\ri}{2} 
\Big( \bar \o  -  \o \Big)\cS_1  ~.
\label{A.9}
\eea

In curved space, the model defined by the action 
$S=\int \rd^4 x\, e \, L_{\rm conf}(\o, \bar \o , \cS , \bar \cS) $ is Weyl-invariant provided 
$\cS$ is inert under the Weyl transformations. 
We can add to \eqref{A.9} a higher-derivative Lagrangian for the dilaton and axion (see, e.g., \cite{Osborn2})
\bea
L &=&  \frac{\k}{2( {\rm Im}\, \cS)^2} \Big[ {\mathfrak D}^2 \cS {\mathfrak D}^2 \bar \cS
- 2 (R^{ab}- \frac 13 \eta^{ab} R) \nabla_a \cS \nabla_b \bar \cS \Big] \non \\
&&+ \frac{1}{12 ( {\rm Im}\, \cS)^4} 
\Big[ \a \nabla^a \cS \nabla_a \cS \nabla^b \bar \cS  \nabla_b \bar \cS
+\b \nabla^a \cS \nabla_a \bar  \cS \nabla^b  \cS  \nabla_b \bar \cS
\Big] 
\label{A.12}
\eea
where 
\bea
{\mathfrak D}^2 \cS := \nabla^a \nabla_a \cS + \frac{\ri}{{\rm Im}\, \cS} \nabla^a \cS \nabla_a \cS ~,
\eea
$\nabla_a$ is the torsion-free Lorentz covariant derivative, 
and $\k, \a$ and $\b$ are numerical parameters. 
The Lagrangian \eqref{A.12} is manifestly invariant under $\sSL(2,{\mathbb R}) $
transformations \eqref{A.10}.
The Weyl invariance follows from the fact that the Fradkin-Tseytlin operator \cite{FT1982}
\bea
\Delta_0 = (\nabla^a \nabla_a)^2 + 2 \nabla^a \big(
	 {R}_{ab} \,\nabla^b 
	- \tfrac{1}{3} {R} \,\nabla_a
	\big)
\eea
is conformal.\footnote{This operator was  re-discovered 
by Paneitz in 1983 \cite{Paneitz} and Riegert in 1984 \cite{Riegert}.}

Within the
$\cN=1$ Poincar\'e supersymmetry,  $\sSL(2,{\mathbb R})$ duality-invariant couplings 
of the dilaton-axion multiplet to general models for self-dual  supersymmetric nonlinear electrodynamics were described in \cite{KT2}, while the case of the supersymmetric 
Born-Infeld action \cite{CF,BG} was first considered in \cite{BMZ}. The results of \cite{KT2} were generalised to supergravity  in  \cite{KMcC}.
  These results can be used 
to couple the $\cN=1$ superconformal $\sU(1)$ duality-invariant model to the
dilaton-axion multiplet described by a chiral scalar $\F$ and its conjugate $\bar \F$.
One obtains
\bea
&&S_{\rm SC}[W,{\bar W}, \F , \bar \F] =
\frac{\ri}{4} \int  \rd^4 x \rd^2 \q  \,\cE \, \F W^2 +{\rm c.c.}
\non \\
&&\quad - \frac{1}{16}  \int \rd^4 x \rd^2 \q \rd^2\bar \q \,E \, (\F-\bar \F)^2 
{W^2\,{\bar W}^2}
\L_{\rm SC}  \left( \frac{\ri}{2} (\F-\bar \F) u,
\frac{\ri}{2} (\F -\bar \F) {\bar u}\right)~,
\label{4.10}
\eea
where $\L_{\rm SC}(u, \bar u) $ is given by eq. \eqref{2.5}.
It is assumed that $\F$ parametrises the lower half-plane, and therefore
$(\F -\bar \F)^{-1}$ exists.\footnote{The axion 
${\mathfrak a}(x) $ and dilaton 
$\vf(x)$ are the component fields of $\F(x,\q)$ defined by $\F|_{\q=0} = {\mathfrak a} -\ri \re^{-\vf}$, which differs from the definition of $\cS$ in  \eqref{43}.} One can add to \eqref{4.10} a higher-derivative super-Weyl and 
 $\sSL (2, {\mathbb R}) $ invariant action for the dilaton-axion multiplet
\cite{K2020}
\bea
S_{\rm DA} [\F, \bar \F]  &=& -\frac{\k}{16}  \int 
\rd^4x \rd^2 \q  \rd^2 \bar \q \, 
E \, \frac{1}{(\F - \bar \F)^2} 
\Big\{ \nabla^2 \F \bar \nabla^2 \bar \F -8 \cD^\a \F G_{\a\ad} \bar \cD^\ad \bar \F\Big\}
\non \\
&& + \frac{\a}{8} \int 
\rd^4x \rd^2 \q  \rd^2 \bar \q \, 
E \, \frac{1}{(\F - \bar \F)^4} \cD^\a \F \cD_\a \F 
\bar \cD_\ad \bar \F \bar \cD^\ad \bar \F~,
\eea
where $\k$ and $\a$ are real parameters, and 
\bea
\nabla^2 \F = \cD^2 \F - 2 \frac{\cD^\a \F \cD_\a \F}{\F -\bar \F} ~, 
\qquad 
\bar \nabla^2 \bar \F = \bar \cD^2 \bar \F 
+ 2 \frac{\bar \cD_\ad \bar \F \bar \cD^\ad \bar  \F}{\F -\bar \F} 
\eea
are $\sSL (2, {\mathbb R}) $ covariant derivatives. 
The complete action $S_{\rm SC}[W,{\bar W}, \F , \bar \F] +S_{\rm DA} [\F, \bar \F] $ has the following fundamental properties:  
(i) it is super-Weyl invariant; and (ii)  it is $\sSL (2, {\mathbb R}) $ duality invariant.
%\\

\noindent
{\bf Acknowledgements:}\\ 
I thank Dmitri Sorokin for kindly informing me of \cite{BLST2}.
I am grateful to Stefan Theisen for a discussion, and to 
Michael Ponds and Emmanouil Raptakis for comments on the manuscript. 
This research was funded in part by the Australian Government through
the Australian 
Research Council, project No. DP200101944.

\appendix 

\section{Conformal duality-invariant electrodynamics} \label{AppendixA} 

In this appendix we re-derive the conformal $\sU(1)$ duality-invariant electrodynamics 
constructed in  \cite{BLST,Kosyakov} following 
the non-supersymmetric approaches described in \cite{KT2} and \cite{IZ1,IZ2}.  

Given a model for nonlinear electrodynamics, 
its Lagrangian $L(F_{ab})$ can be expressed in terms of the 
 two independent invariants of the electromagnetic field,
\bea
\a = \frac{1}{4} \, F^{ab} F_{ab}~, \qquad 
\b = \frac{1}{4} \, F^{ab} \tilde{F}_{ab} ~.
\eea
Introducing the complex combination $\o = \a + {\rm i} \, \b$, the Lagrangian 
can be rewritten in the form 
\bea
L(\o, \bar \o)  = -\hf \, \Big( \o + \bar{\o} \Big) +
\o \, \bar{\o} \; \L (\o, \bar{\o} )~.
\label{A.2}
\eea
Then, the GZGR condition for $\sU(1)$ duality invariance is given by 
\bea
{\rm Im} \bigg\{ \frac{\pa (\o \, \L) }{\pa \o}
- \bar{\o}\,
\left( \frac{\pa (\o \, \L )  }{\pa \o} \right)^2 \bigg\} = 0~,
\label{GZ4}
\eea
see \cite{KT2} for the technical details.

Requiring the action functional to be conformally invariant constrains $\L(\o,\bar \o)$ 
 in \eqref{A.2} to have the following functional form 
 \bea
{\bm \L}_{\rm conf}(\o, \bar \o) = (\o\bar \o)^{-\hf} f(\o/\bar \o)~, 
\qquad f: S^1 \to {\mathbb R}~.
\eea
Only a restricted class of such models prove to be compatible with $\sU(1)$ duality invariance,
specifically
 \bea
 \L_{\rm conf}(\o, \bar \o) = \frac{y }{\sqrt{\o\bar \o} } 
- x \Big( \frac{1}{\o} + \frac{1}{\bar \o}\Big)~,
\eea
with $x$ and $ y $ being real coefficients. These coefficients turn out to be constrained by  requiring the self-duality equation \eqref{GZ4} to hold, specifically
% leads to 
\bea
\L_{\rm conf}(\o, \bar \o) = \frac{\sinh \g }{\sqrt{\o\bar \o} } 
- \hf (\cosh \g -1) \Big( \frac{1}{\o} + \frac{1}{\bar \o}\Big)~,
\eea
with $\g$ a real parameter. The resulting Lagrangian is 
\bea
L_{\rm conf}(\o, \bar \o) = - \hf \cosh \g  \Big( {\o} + {\bar \o}\Big)
+ {\sinh \g }{\sqrt{\o\bar \o} } ~,
\label{A.6}
\eea
which is exactly the model proposed in  \cite{BLST} (see also \cite{Kosyakov}). It was called  ``ModMax electrodynamics'' in \cite{BLST}.

It is instructive to re-derive the conformal duality-invariant model using 
the IZ approach \cite{IZ1,IZ2}. Their reformulation of nonlinear 
electrodynamics is obtained by replacing $L(F_{ab}) \to \widetilde{L}(F_{ab} , V_{ab})$, where  $V_{ab}=-V_{ba}$ is an auxiliary unconstrained bivector, which is equivalent to a pair of symmetric rank-2 spinors,
$V_{\a\b}= V_{\b\a}$ and its conjugate $\bar  V_{\ad\bd}$.
The new Lagrangian $\widetilde{L}$ is at most quadratic with respect to
the electromagnetic field strength $F_{ab}$, while the self-interaction is described 
by a nonlinear function of the auxiliary variables, $L_{\rm int} (V_{ab})$,
\bea
 \widetilde{L}(F_{ab} , V_{ab}) = \frac{1}{4} F^{ab}F_{ab} +\hf V^{ab}V_{ab} 
 - V^{ab}F_{ab} + L_{\rm int} (V_{ab})~.
\eea 
The original theory  $L(F_{ab})$ is obtained from 
$\widetilde{L}(F_{ab} , V_{ab})$ by integrating out the auxiliary variables. 
In terms of $\widetilde{L}(F_{ab} , V_{ab})$, the condition of $\sU(1)$ duality invariance 
was shown \cite{IZ1,IZ2}  to be equivalent to the requirement that the self-interaction
\bea
L_{\rm int} (V_{ab}) = L_{\rm int} (\n, \bar \n)~, \qquad \n:=V^{\a\b}V_{\a\b}
\eea
is invariant under linear $\sU(1)$  transformations $\n \to \re^{\ri \vf} \n$, with $\vf \in \mathbb R$, therefore
\bea
L_{\rm int} (\n, \bar \n)= f (\n \bar \n)~.
\eea

A unique conformal duality-invariant model corresponds to the choice
\bea
L_{\rm int, conf} = \k \sqrt{\n \bar \n}~,
\eea
with $\x$ a coupling constant. Integrating out the auxiliary variables lead to the model 
\eqref{A.6}, in which 
\bea
\sinh \g = \frac{\k}{1-(\k/2)^2} ~.
\eea

%%%%%%%%%%%%%%%%%%%%%%%%%%%%%%%%%
%%%%%%%%%%%%%%%%%%%%%%%%%%%%%%%%

\section{Elements of $\cN = 1$ supergravity} \label{AppendixB}

General $\cN=1$ supergravity-matter systems can be naturally described in superspace using 
the Grimm-Wess-Zumino geometry \cite{GWZ}
in conjunction with the super-Weyl transformations discovered  in  \cite{HT}.
 The structure group in this setting is $\sSL(2, \dsC) $, and the geometry of curved superspace is described by the covariant derivatives
\bea
\cD_{A}=(\cD_{{a}}, \cD_{{\a}},\cDB^\ad)
=E_{A}{}^M \pa_M
+ \Omega_A{}^{\b\g} M_{\b\g} 
  + \bar{\Omega}_A{}^{ \dot{\b} \dot{\g} } \bar M_{\dot{\b}\dot{\g}}~,  
\label{CovDev}
\eea
where $M_{\b\g} =M_{\g\b}$  and ${\bar M}_{\bd \gd} ={\bar M}_{\gd \bd}$  are the Lorentz generators.
The covariant derivatives  obey the graded commutation relations 
(see \cite{Ideas} for the  derivation)
\begin{subequations}\label{algebra}
\bea
& \{ \cD_\a , {\bar \cD}_\ad \} = -2{\rm i} \cD_{\a \ad} ~,\\
&
\{\cD_\a, \cD_\b \} = -4{\bar R} M_{\a \b}~,
 \qquad
\{ {\bar \cD}_\ad, {\bar \cD}_\bd \} =  4R {\bar M}_{\ad \bd}~, 
 \\
&\left[ \cD_{\a} , \cD_{ \b \bd } \right]
      = 
     {\rm i}  {\ve}_{\a \b}
\Big({\bar R}\,\cDB_\bd + G^\g{}_\bd \cD_\g
- \cD^\g G^\d{}_\bd  M_{\g \d}
+2{\bar W}_\bd{}^{\gd \dot{\d}}
{\bar M}_{\gd \dot{\d} }  \Big)
+ {\rm i} \cDB_\bd {\bar R} \, M_{\a \b}~,
~~~~~~\\
&\left[ { \bar \cD}_{\ad} , \cD_{ \b \bd } \right]
      =  -{\rm i}{\ve}_{\ad \bd}
\Big(R\,\cD_\b + G_\b{}^{\dot{\g}}  \cDB_{\dot{\g}}
-\cDB^\gd G_\b{}^{\dot{\d}}
{\bar M}_{\gd \dot{\d}}
+2W_\b{}^{\g \d}
M_{\g \d} \Big)
- {\rm i} \cD_\b R  {\bar M}_{\ad \bd}~.~~~~~~~ ~~
\eea
\esubeq
Here the torsion tensors $R$, $G_a = {\bar G}_a$ and
$W_{\a \b \g} = W_{(\a \b\g)}$ satisfy the  Bianchi identities:
\begin{subequations}
\bea
&\cDB_\ad R= 0~,~~~~~~\cDB_\ad W_{\a \b \g} = 0~,
\\
&
\cDB^\gd G_{\a \gd} = \cD_\a R~,~~~~~~
\cD^\g W_{\a \b \g} = {\rm i} \,\cD_{(\a }{}^\gd G_{\b) \gd}~.
\eea
\end{subequations}

An infinitesimal super-Weyl transformation of the covariant derivatives \cite{HT} 
\begin{subequations} 
\label{superweyl}
\bea
\d_\s \cD_\a &=& ( {\bar \s} - \hf \s)  \cD_\a + \cD^\b \s \, M_{\a \b}  ~, \\
\d_\s \bar \cD_\ad & = & (  \s -  \hf {\bar \s})
\bar \cD_\ad +  ( \bar \cD^\bd  {\bar \s} )  {\bar M}_{\ad \bd} 
\eea
\end{subequations}
induces
the following variations of the torsion superfields
\begin{subequations} 
\bea
\d_\s R &=& 2\s R +\frac{1}{4} (\bar \cD^2 -4R ) \bar \s ~, \\
\d_\s G_{\a\ad} &=& \hf (\s +\bar \s) G_{\a\ad} +\ri \cD_{\a\ad} ( \s- \bar \s) ~, 
\label{s-WeylG}\\
\d_\s W_{\a\b\g} &=&\frac{3}{2} \s W_{\a\b\g}~.
\label{s-WeylW}
\eea
\end{subequations} 
Here the super-Weyl parameter $\s$ is a covariantly chiral scalar superfield,  $\bar \cD_\ad \s =0$. 
The super-Weyl transformations belong to the  
gauge group of conformal supergravity.

Integrals over the full superspace and its chiral subspace are related as follows:
\bea
\int \rd^4 x \,{\rm d}^2\q \,{\rm d}^2{\bar \q}\,E\, U 
= -\frac 14 \int \rd^4 x \,{\rm d}^2\q \,
\,\cE\, (\bar\cD^2 -4R) U 
~, \qquad E^{-1}= {\rm Ber}(E_A{}^M)~,
\eea
where $\cE$ is the chiral density (see, e.g., \cite{Ideas} for the derivation). The super-Weyl transformation laws of the integration measures are
\bea
\d_\s E = - (\s+\bar \s) E~, \qquad \d_\s \cE = - 3\s \cE~.
\eea

%%%%%%%%%%%%%%%%%%%%%%%%%%%%%%%
%%%%%%%%%%%%%%%%%%%%%%%%%%%%%%%

\section{Elements of $\cN = 2$ supergravity} \label{AppendixC} 

In this appendix we give a summary of the so-called $\sSU(2)$ superspace, the  off-shell formulation for $\cN=2$ conformal supergravity
developed in \cite{KLRT-M}.
The structure group in this setting is $\sSL(2, \dsC) \times \sSU(2)$, 
and the covariant derivatives are
\bea
\cD_A = (\cD_a, \cD_\a^i, \bar \cD^\ad_i)
= E_A{}^M \pa_M  + \Omega_A{}^{\b\g} M_{\b\g} 
  + \bar{\Omega}_A{}^{ \dot{\b} \dot{\g} } \bar M_{\dot{\b}\dot{\g}}
  + \Phi_A{}^{kl} J_{kl}
  ~.
  \label{C.1}
\eea
Here $J_{kl}$ are the generators of the group $\sSU(2)$.
The spinor covariant derivatives obey the anti-commutation relations 
\begin{subequations} 
\bea
\{\cD_\a^i,\cD_\b^j\}&=&\phantom{+}
4S^{ij}M_{\a\b}
+2\ve^{ij}\ve_{\a\b}Y^{\g\d}M_{\g\d}
+2\ve^{ij}\ve_{\a\b}\bar{W}^{\gd\dd}\bar{M}_{\gd\dd}
\non\\
&&
+2 \ve_{\a\b}\ve^{ij}S^{kl}J_{kl}
+4 Y_{\a\b}J^{ij}~,
\label{acr1} \\
\{\cDB^\ad_i,\cDB^\bd_j\}&=&
-4\bar{S}_{ij}\bar{M}^{\ad\bd}
-2\ve_{ij}\ve^{\ad\bd}\bar{Y}^{\gd\dd}\bar{M}_{\gd\dd}
-2\ve_{ij}\ve^{\ad\bd}{W}^{\g\d}M_{\g\d} \non \\
&&
-2\ve_{ij}\ve^{\ad\bd}\bar{S}^{kl}J_{kl}
-4\bar{Y}^{\ad\bd}J_{ij}~,
\label{acr2} \\
\{\cD_\a^i,\cDB^\bd_j\}&=&
-2\ri\d^i_j(\s^c)_\a{}^\bd\cD_c
+4\d^{i}_{j}G^{\d\bd}M_{\a\d}
+4\d^{i}_{j}G_{\a\gd}\bar{M}^{\gd\bd}
+8 G_\a{}^\bd J^{i}{}_{j}~.
\label{acr3}
\eea
\end{subequations}
Here the real four-vector $G_{\a \ad} $,
the complex symmetric  tensors $S^{ij}=S^{ji}$, $W_{\a\b}=W_{\b\a}$, 
$Y_{\a\b}=Y_{\b\a}$ and their complex conjugates 
$\bar{S}_{ij}:=\overline{S^{ij}}$, $\bar{W}_{\ad\bd}:=\overline{W_{\a\b}}$,
$\bar{Y}_{\ad\bd}:=\overline{Y_{\a\b}}$ obey additional differential constraints implied 
by the Bianchi identities and given in \cite{KLRT-M}.

In $\sSU(2)$ superspace, the gauge group of conformal supergravity includes super-Weyl transformations. 
An infinitesimal  super-Weyl transformation of $\cD_A$ \cite{KLRT-M} is
\begin{subequations} \label{super-Weyl} 
\bea
\d_{\s} \cD_\a^i&=&\hf\sba\cD_\a^i+(\cD^{\g i}\s)M_{\g\a}-(\cD_{\a k}\s)J^{ki}~, \label{super-Weyl1} \\
\d_{\s} \cDB_{\ad i}&=&\hf\s\cDB_{\ad i}+(\cDB^{\gd}_{i}\sba)\bar{M}_{\gd\ad}
+(\cDB_{\ad}^{k}\sba)J_{ki}~, 
\label{super-Weyl2} 
\eea
\end{subequations}
where  the parameter $\s$ is an arbitrary covariantly chiral superfield, $\bar \cD^\ad_i \s =0$.
The  dimension-1 torsion superfields transform 
 as follows:
\begin{subequations}
\bea
\d_{\s} S^{ij}&=&\sba S^{ij}-{\frac14}\cD^{\g(i}\cD^{j)}_\g \s~, 
\label{super-Weyl-S} \\
\d_{\s} Y_{\a\b}&=&\sba Y_{\a\b}-{\frac14}\cD^{k}_{(\a}\cD_{\b)k}\s~,
\label{super-Weyl-Y} \\
\d_{\s} G_{\a\bd} &=&
\hf(\s+\sba)G_{\a\bd} -{\frac{\ri}4}
\cD_{\a \bd} (\s-\sba)~, \\
\d_{\s} {W}_{\a \b}&=&\s {W}_{\a \b }~. 
\label{C4.d}
\eea
\end{subequations}
Eq. \eqref{C4.d} shows that $W_{\a\b}$ is the super-Weyl tensor. 

The covariantly chiral projecting operator \cite{Muller,KT-M} is 
\bea
\bar{\D}
&=&\frac{1}{96} \Big((\cDB^{ij}+16\bar{S}^{ij})\cDB_{ij}
-(\cDB^{\ad\bd}-16\bar{Y}^{\ad\bd})\cDB_{\ad\bd} \Big)
\non\\
&=&\frac{1}{96} \Big(\cDB_{ij}(\cDB^{ij}+16\bar{S}^{ij})
-\cDB_{\ad\bd}(\cDB^{\ad\bd}-16\bar{Y}^{\ad\bd}) \Big)~,
\label{chiral-pr}
\eea
where $\cDB^{\ad\bd}:=\cDB^{(\ad}_k\cDB^{\bd)k}$. 
The  fundamental property  of $\bar \D$ is that $\bar{\D} U$ is covariantly chiral,
for any scalar and isoscalar superfield $U$,
that is ${\bar \cD}^{\ad}_i \bar{\D} U =0$. 
For any super-Weyl inert scalar $U$ it holds that 
\bea
\d_\s U = 0 \quad \Longrightarrow \quad 
\d_\s \bar \D U &=& 2\s \bar \D U~, 
\eea
The operator $\bar \D$ relates an integral over the full superspace to that over its chiral subspace:
\bea
\int \rd^4 x \,{\rm d}^4\q \,{\rm d}^4{\bar \q}\,E\, U 
=  \int \rd^4 x \,{\rm d}^4\q \,
\,\cE\, \bar\D U 
~, \qquad E^{-1}= {\rm Ber}(E_A{}^M)~,
\label{chiral_action_rule}
\eea
with $\cE$ the chiral density.\footnote{A   derivation of \eqref{chiral_action_rule} is given \cite{KT-M}.} 
The super-Weyl transformation laws of the integration measures are
\bea
\d_\s E = 0~, \qquad \d_\s \cE = - 2\s \cE~.
\label{C.8}
\eea

The superfield $\X$ in \eqref{3.5} is the following composite scalar \cite{BdeWKL}:
\bea
\X :=  \frac{1}{6}  \bar{\cD}^{ij} \bar S_{ij}+  \bar S^{ij} \bar S_{ij}
+ \bar Y_{\dalpha \dbeta} \bar Y^{\dalpha \dbeta}~,
\label{C.7}
\eea   
The fundamental properties of $\X$ are as follows \cite{BdeWKL}: \\
(i) $\X$ is covariantly chiral, 
\begin{subequations}
\bea
\bar \cD^\ad_i \X=0~;
\eea
 (ii) the super-Weyl varation of $\X$  is 
 \bea
 \d_\s \X = 2\s \X -2 \bar \D \bar  \s ~;
\label{sW-X}
 \eea
 \end{subequations}
(iii) the functional \eqref{3.20a}
is a topological invariant
related to the difference 
of the Gauss-Bonnet and Pontryagin invariants.

%%%%%%%%%%%%%%%%%%%%%%%%%%%%%%%%%%%
%%%%%%%%%%%%%%%%%%%%%%%%%%%%%%%%%%%

\begin{footnotesize}

\end{footnotesize}

%%%%%%%%%%%%%%%%%%%%%%%%%%%%%%%%%%%%
%%%%%%%%%%%%%%%%%%%%%%%%%%%%%%%%%%%%


\begin{thebibliography}{66}


\bibitem{MO}
C.~Montonen and D.~I.~Olive,
``Magnetic monopoles as gauge particles?,''
Phys.\ Lett.\ B {\bf 72}, 117 (1977).

\bibitem{Osborn}
H.~Osborn,
``Topological charges for N=4 supersymmetric gauge 
theories and monopoles of spin 1,''
Phys.\ Lett.\ B {\bf 83},  321 (1979).

\bibitem{Sen}
A.~Sen,
``Dyon-monopole bound states, self-dual harmonic 
forms  on the multi-monopole moduli space, 
and SL(2,Z) invariance in string theory,''
Phys.\ Lett.\ B {\bf 329}, 217 (1994)
[arXiv:hep-th/9402032].



\bibitem{SW1}
N.~Seiberg and E.~Witten,
``Electric - magnetic duality, monopole condensation, 
and confinement in N=2
supersymmetric Yang-Mills theory,''
Nucl.\ Phys.\ B {\bf 426},  19 (1994)
(Erratum-ibid.\ B {\bf 430}, 485  (1994))
[arXiv:hep-th/9407087].

\bibitem{SW2}
N.~Seiberg and E.~Witten,
``Monopoles, duality and chiral symmetry breaking in 
N=2 supersymmetric QCD,''
Nucl.\ Phys.\ B {\bf 431}, 484 (1994) 
[arXiv:hep-th/9408099].

\bibitem{Maldacena}
J.~M.~Maldacena,
``The large N limit of superconformal field theories 
and supergravity,''
Adv.\ Theor.\ Math.\ Phys.\  {\bf 2}, 231 (1998) 
[arXiv:hep-th/9711200].


\bibitem{GKPR}
F.~Gonzalez-Rey, B.~Kulik, I.~Y.~Park and M.~Ro\v{c}ek,
``Self-dual effective action of N = 4 super-Yang-Mills,''
Nucl.\ Phys.\ B {\bf 544}, 218 (1999) 
[arXiv:hep-th/9810152].

\bibitem{KT1}
S.~M.~Kuzenko and S.~Theisen,
``Supersymmetric duality rotations,''
JHEP {\bf 0003}, 034 (2000)
[arXiv:hep-th/0001068].

\bibitem{Schwarz:2013wra}
J.~H.~Schwarz,
``Highly effective actions,''
JHEP \textbf{01}, 088 (2014)
%doi:10.1007/JHEP01(2014)088
[arXiv:1311.0305 [hep-th]].

\bibitem{PvU}
V.~Periwal and R.~von Unge,
``Accelerating D-branes,''
Phys.\ Lett.\ B {\bf 430}, 71 (1998) 
[arXiv:hep-th/9801121].

\bibitem{G-RR}
F.~Gonzalez-Rey and M.~Ro\v{c}ek,
``Nonholomorphic N = 2 terms in N = 4 SYM: 
1-loop calculation in N = 2 superspace,''
Phys.\ Lett.\ B {\bf 434},  303(1998) 
[arXiv:hep-th/9804010].

\bibitem{BK98}
I.~L.~Buchbinder and S.~M.~Kuzenko,
``Comments on the background field method in harmonic
superspace: Non-holomorphic corrections in N = 4 SYM,''
Mod.\ Phys.\ Lett.\ A {\bf 13}, 1623 (1998) 
[arXiv:hep-th/9804168].

\bibitem{BBK}
E.~I.~Buchbinder, I.~L.~Buchbinder and S.~M.~Kuzenko,
``Non-holomorphic effective potential in N = 4 SU(n) SYM,''
Phys.\ Lett.\ B {\bf 446}, 216  (1999)
[arXiv:hep-th/9810239].

\bibitem{LvU}
D.~A.~Lowe and R.~von Unge,
``Constraints on higher derivative operators 
in maximally supersymmetric  gauge theory,''
JHEP {\bf 9811}, 014 (1998) 
[arXiv:hep-th/9811017].
  


\bibitem{K2004}
S.~M.~Kuzenko,
``Self-dual effective action of N = 4 SYM revisited,''
JHEP \textbf{03}, 008 (2005)
%doi:10.1088/1126-6708/2005/03/008
[arXiv:hep-th/0410128 [hep-th]]. 

\bibitem{BPT}
I.~L.~Buchbinder, A.~Y.~Petrov and A.~A.~Tseytlin,
``Two-loop N=4 superYang-Mills effective action and interaction between D3-branes,''
Nucl. Phys. B \textbf{621}, 179-207 (2002)
%doi:10.1016/S0550-3213(01)00575-2
[arXiv:hep-th/0110173 [hep-th]].

\bibitem{KT2}
S.~M.~Kuzenko and S.~Theisen,
``Nonlinear self-duality and supersymmetry,''
Fortsch.\ Phys.\  {\bf 49}, 273 (2001) [arXiv:hep-th/0007231].



\bibitem{GZ1}
M. K.~Gaillard and B.~Zumino,
``Duality rotations for interacting fields,''
Nucl.\ Phys.\  {\bf B193},  221 (1981). 
%%CITATION = NUPHA,B193,221;%%

\bibitem{GR1}
G.~W.~Gibbons and D.~A.~Rasheed,
``Electric-magnetic duality rotations in nonlinear electrodynamics,''
Nucl.\ Phys.\  {\bf B454}, 185 (1995) 
[arXiv:hep-th/9506035].
%%CITATION = HEP-TH 9506035;%%

\bibitem{GR2}
G. W.~Gibbons and D. A.~Rasheed,
``SL(2,R) invariance of non-linear electrodynamics
coupled to an axion and a dilaton,''
Phys.\ Lett.\  {\bf B365}, 46 (1996) 
[hep-th/9509141].
%%CITATION = HEP-TH 9509141;%%

\bi{GZ2}
M.~K.~Gaillard and B.~Zumino,
``Self-duality in nonlinear electromagnetism,''
in {\it Supersymmetry and Quantum Field Theory},
J.~Wess and V.~P.~Akulov (Eds.), Springer Verlag, 1998, pp. 121--129 [arXiv:hep-th/9705226].
%%CITATION = HEP-TH 9705226;%%

\bi{GZ3}
M.~K.~Gaillard and B.~Zumino,
``Nonlinear electromagnetic self-duality
and Legendre transformations,'' in {\it Duality and
Supersymmetric Theories}, D.~I.~Olive and
P.~C.~West (Eds.), Cambridge University Press,
1999, pp. 33--48 [hep-th/9712103].
%%CITATION = HEP-TH 9712103;%%


\bibitem{IZ_N3} 
  E.~A.~Ivanov and B.~M.~Zupnik,
  ``N=3 supersymmetric Born-Infeld theory,''
  Nucl.\ Phys.\ B {\bf 618}, 3 (2001)
  %doi:10.1016/S0550-3213(01)00540-5
  [hep-th/0110074].



%\cite{Ivanov:2002ab}
\bibitem{IZ1} 
  E.~A.~Ivanov and B.~M.~Zupnik,
  ``New representation for Lagrangians of self-dual nonlinear electrodynamics,''
 in {\it Supersymmetries and Quantum Symmetries. Proceedings of the 16th Max Born Symposium, SQS'01: Karpacz, Poland, September 21--25, 2001}, E. Ivanov (Ed.), Dubna, 2002, pp. 235--250 
 [hep-th/0202203].
  %%CITATION = HEP-TH/0202203;%%

%\cite{Ivanov:2003uj}
\bibitem{IZ2} 
  E.~A.~Ivanov and B.~M.~Zupnik,
  ``New approach to nonlinear electrodynamics: Dualities as symmetries of interaction,''
  Phys.\ Atom.\ Nucl.\  {\bf 67}, 2188 (2004)
  [Yad.\ Fiz.\  {\bf 67}, 2212 (2004)]
  [hep-th/0303192].
  %%CITATION = HEP-TH/0303192;%%



\bibitem{IZ3}
E.~A.~Ivanov and B.~M.~Zupnik,
``Bispinor auxiliary fields in duality-invariant electrodynamics revisited,''
Phys. Rev. D \textbf{87}, no.6, 065023 (2013)
%doi:10.1103/PhysRevD.87.065023
[arXiv:1212.6637 [hep-th]].


%\cite{Aschieri:2008ns}
\bibitem{AFZ}
P.~Aschieri, S.~Ferrara and B.~Zumino,
``Duality rotations in nonlinear electrodynamics and in extended supergravity,''
  Riv.\ Nuovo Cim.\  {\bf 31}, 625 (2008)
  [arXiv:0807.4039 [hep-th]].
  %%CITATION = RNCIB,031,625;%%


%\cite{Chemissany:2011yv}
\bibitem{Chemissany:2011yv} 
  W.~Chemissany, R.~Kallosh and T.~Ortin,
  ``Born-Infeld with higher derivatives,''
  Phys.\ Rev.\ D {\bf 85}, 046002 (2012)
  [arXiv:1112.0332 [hep-th]].
  %%CITATION = ARXIV:1112.0332;%%

\bibitem{AF} 
  P.~Aschieri and S.~Ferrara,
  ``Constitutive relations and Schroedinger's formulation of nonlinear electromagnetic theories,''
  JHEP {\bf 1305}, 087 (2013)
  %doi:10.1007/JHEP05(2013)087
  [arXiv:1302.4737 [hep-th]].

\bibitem{AFT} 
  P.~Aschieri, S.~Ferrara and S.~Theisen,
  ``Constitutive relations, off shell duality rotations and the hypergeometric form of Born-Infeld theory,''
  Springer Proc.\ Phys.\  {\bf 153}, 23 (2014)
%  doi:10.1007/978-3-319-03774-5_2
  [arXiv:1310.2803 [hep-th]].


\bibitem{Tanii}
Y.~Tanii,
{\it Introduction to supergravities in diverse dimensions},
hep-th/9802138.
%%CITATION = HEP-TH 9802138;%%

\bibitem{AT}
M.~Araki and Y.~Tanii,
``Duality symmetries in non-linear gauge theories,''
Int.\ J.\ Mod.\ Phys.\  {\bf A14}, 1139 (1999) 
[hep-th/9808029].
%%CITATION = HEP-TH 9808029;%%

\bibitem{ABBZ} 
  P.~Aschieri, D.~Brace, B.~Morariu and B.~Zumino,
  ``Nonlinear self-duality in even dimensions,''
  Nucl.\ Phys.\ B {\bf 574}, 551 (2000)
%  doi:10.1016/S0550-3213(00)00019-5
  [hep-th/9909021].
  

\bibitem{Tanii2} Y. Tanii, {\it Introduction to Supergravity}, Springer, 2014.

 
%\cite{Kuzenko:2005wh}
\bibitem{KMcC}
S.~M.~Kuzenko and S.~A.~McCarthy,
``Nonlinear self-duality and supergravity,''
JHEP {\bf 0302}, 038 (2003)
[hep-th/0212039].  

\bibitem{KMcC2}
S.~M.~Kuzenko and S.~A.~McCarthy,
``On the component structure of N=1 supersymmetric nonlinear electrodynamics,''
JHEP \textbf{05}, 012 (2005)
%doi:10.1088/1126-6708/2005/05/012
[arXiv:hep-th/0501172 [hep-th]].

%\cite{Kuzenko:2012ht}
\bibitem{K12} 
  S.~M.~Kuzenko,
  ``Nonlinear self-duality in N=2 supergravity,''
  JHEP {\bf 1206}, 012 (2012)
  [arXiv:1202.0126 [hep-th]].
  %%CITATION = ARXIV:1202.0126;%%



\bibitem{Ketov}
S.~V.~Ketov,
``A manifestly N=2 supersymmetric Born-Infeld action,''
Mod. Phys. Lett. A \textbf{14}, 501 (1999)
%doi:10.1142/S0217732399000559
[arXiv:hep-th/9809121 [hep-th]];
``Born-Infeld-Goldstone superfield actions for gauge-fixed D5- and D3-branes in 6d,''
Nucl.\ Phys.\  {\bf B553}, 250 (1999) 
[hep-th/9812051].

\bibitem{BIK1}
S.~Bellucci, E.~Ivanov and S.~Krivonos,
``N=2 and N=4 supersymmetric Born-Infeld theories from nonlinear realizations,''
Phys. Lett. B \textbf{502}, 279 (2001)
%doi:10.1016/S0370-2693(01)00142-3
[arXiv:hep-th/0012236 [hep-th]].


\bibitem{BIK2}
S.~Bellucci, E.~Ivanov and S.~Krivonos,
``Towards the complete N = 2 superfield Born-Infeld 
action with partially broken N = 4 supersymmetry,''
Phys.\ Rev.\ D {\bf 64},    025014 (2001) 
[arXiv:hep-th/0101195].

%\cite{Broedel:2012gf}
\bibitem{BCFKR} 
  J.~Broedel, J.~J.~M.~Carrasco, S.~Ferrara, R.~Kallosh and R.~Roiban,
  ``N=2 supersymmetry and U(1)-duality,''
  Phys.\ Rev.\ D {\bf 85}, 125036 (2012)
  [arXiv:1202.0014 [hep-th]].
  %%CITATION = ARXIV:1202.0014;%%

\bibitem{CK}
J.~J.~M.~Carrasco and R.~Kallosh,
``Hidden supersymmetry may imply duality invariance,''
[arXiv:1303.5663 [hep-th]].

\bibitem{IZ4}
E.~A.~Ivanov and B.~M.~Zupnik,
``Self-dual $\mathcal N=2$ Born-Infeld theory through auxiliary superfields,''
JHEP \textbf{05}, 061 (2014)
%doi:10.1007/JHEP05(2014)061
[arXiv:1312.5687 [hep-th]].


\bibitem{K13} 
  S.~M.~Kuzenko,
  ``Duality rotations in supersymmetric nonlinear electrodynamics revisited,''
  JHEP {\bf 1303}, 153 (2013)
 % doi:10.1007/JHEP03(2013)153
  [arXiv:1301.5194 [hep-th]].



\bibitem{ILZ}
E.~Ivanov, O.~Lechtenfeld and B.~Zupnik,
``Auxiliary superfields in N=1 supersymmetric self-dual electrodynamics,''
JHEP \textbf{05}, 133 (2013)
%doi:10.1007/JHEP05(2013)133
[arXiv:1303.5962 [hep-th]].

\bibitem{K19duality}
S.~M.~Kuzenko,
``Manifestly duality-invariant interactions in diverse dimensions,''
Phys. Lett. B \textbf{798}, 134995 (2019)
%doi:10.1016/j.physletb.2019.134995
[arXiv:1908.04120 [hep-th]].

 \bibitem{Novotny} 
  J.~Novotn\'y,
  ``Self-duality, helicity conservation and normal ordering in nonlinear QED,''
  Phys.\ Rev.\ D {\bf 98}, no. 8, 085015 (2018)
  %doi:10.1103/PhysRevD.98.085015
  [arXiv:1806.02167 [hep-th]].


\bibitem{INZ} 
  E.~A.~Ivanov, A.~J.~Nurmagambetov and B.~M.~Zupnik,
  ``Unifying the PST and the auxiliary tensor field formulations of 4D self-duality,''
  Phys.\ Lett.\ B {\bf 731}, 298 (2014)
  %doi:10.1016/j.physletb.2014.02.052
  [arXiv:1401.7834 [hep-th]].


\bibitem{PST1} 
  P.~Pasti, D.~P.~Sorokin and M.~Tonin,
  ``Duality symmetric actions with manifest space-time symmetries,''
  Phys.\ Rev.\ D {\bf 52}, 4277 (1995)
  [hep-th/9506109].
  %%CITATION = HEP-TH/9506109;%%

%\cite{Pasti:2012wv}
\bibitem{PST2} 
  P.~Pasti, D.~Sorokin and M.~Tonin,
  ``Covariant actions for models with nonlinear twisted self-duality,''
  Phys.\ Rev.\ D {\bf 86}, 045013 (2012)
  [arXiv:1205.4243 [hep-th]].
  %%CITATION = ARXIV:1205.4243;%%

\bibitem{BdeWKL}
D.~Butter, B.~de Wit, S.~M.~Kuzenko and I.~Lodato,
``New higher-derivative invariants in N=2 supergravity and the Gauss-Bonnet term,''
JHEP \textbf{12}, 062 (2013)
%doi:10.1007/JHEP12(2013)062
[arXiv:1307.6546 [hep-th]].

\bibitem{K13anomaly}
S.~M.~Kuzenko,
``Super-Weyl anomalies in N=2 supergravity and (non)local effective actions,''
JHEP {\bf 1310}  (2013) 151
%  doi:10.1007/JHEP10(2013)151
[arXiv:1307.7586 [hep-th]].


\bibitem{BLST}
I.~Bandos, K.~Lechner, D.~Sorokin and P.~K.~Townsend,
``A non-linear duality-invariant conformal extension of Maxwell's equations,''
Phys. Rev. D \textbf{102}, 121703 (2020)
%doi:10.1103/PhysRevD.102.121703
[arXiv:2007.09092 [hep-th]].

\bibitem{Kosyakov}
B.~P.~Kosyakov,
``Nonlinear electrodynamics with the maximum allowable symmetries,''
Phys. Lett. B \textbf{810}, 135840 (2020)
%doi:10.1016/j.physletb.2020.135840
[arXiv:2007.13878 [hep-th]].


\bibitem{BLST2}
I.~Bandos,  K.~Lechner, D.~Sorokin and P.~K.~Townsend, ``ModMax meets Susy,'' 
 [arXiv:2106.07547 [hep-th]].



\bibitem{WB} J.~Wess and J.~Bagger,
{\it Supersymmetry and Supergravity},
Princeton University Press, Princeton, 1992.


 \bibitem{Ideas} 
I.~L. Buchbinder and S.~M. Kuzenko, {\it Ideas and Methods of Supersymmetry and
Supergravity, Or a Walk Through Superspace},
 IOP, Bristol, 1995 (Revised Edition 1998).
   


\bibitem{FZ} 
  S.~Ferrara and B.~Zumino,
  ``Supergauge invariant Yang-Mills theories,''
  Nucl.\ Phys.\ B {\bf 79}, 413 (1974).

\bibitem{WZ}
J.~Wess and B.~Zumino,
 ``Superfield Lagrangian for supergravity,''
 Phys.\ Lett.\  B {\bf 74}, 51 (1978).
%%CITATION = PHLTA,B74,51;%%

\bibitem{K19}
S.~M.~Kuzenko,
``Superconformal vector multiplet self-couplings and generalised Fayet-Iliopoulos terms,''
Phys. Lett. B \textbf{795}, 37-41 (2019)
%doi:10.1016/j.physletb.2019.05.047
[arXiv:1904.05201 [hep-th]].



\bibitem{CF}
S.~Cecotti and S.~Ferrara,
``Supersymmetric Born-Infeld Lagrangians,''
Phys.\ Lett.\ B {\bf 187}, 335 (1987).

\bibitem{BG}
J.~Bagger and A.~Galperin,
``A new Goldstone multiplet for partially 
broken supersymmetry,''
Phys.\ Rev.\ D {\bf 55}, 1091 (1997) 
[arXiv:hep-th/9608177].

\bibitem{RT}
M.~Ro\v{c}ek and A.~A.~Tseytlin,
``Partial breaking of global D = 4 supersymmetry, 
constrained  superfields, and 3-brane actions,''
Phys.\ Rev.\ D {\bf 59},  106001 (1999) 
[arXiv:hep-th/9811232].



\bibitem{KT-M16} 
  S.~M.~Kuzenko and G.~Tartaglino-Mazzucchelli,
  ``Nilpotent chiral superfield in N=2 supergravity and 
  partial rigid supersymmetry breaking,''
  JHEP {\bf 1603}, 092 (2016)
  %doi:10.1007/JHEP03(2016)092
  [arXiv:1512.01964 [hep-th]].


\bibitem{KakuT}
M.~Kaku and P.~K.~Townsend,
``Poincar\'e supergravity as broken superconformal gravity,''
Phys.\ Lett.\  B {\bf 76},  54 (1978).
%%CITATION = PHLTA,B76,54;%%  


\bibitem{FGKV}
S.~Ferrara, L.~Girardello, T.~Kugo and A.~Van Proeyen,
``Relation between different auxiliary field formulations of N=1 supergravity
coupled to matter,''   Nucl.\ Phys.\  B {\bf 223},   191 (1983).


\bibitem{K15Corfu} 
S.~M.~Kuzenko,
``Supersymmetric spacetimes from curved superspace,''
PoS CORFU {\bf 2014}, 140 (2015)
%doi:10.22323/1.231.0140
[arXiv:1504.08114 [hep-th]].
  

%\cite{Grimm:1977xp}
\bibitem{GSW}
 R.~Grimm, M.~Sohnius and J.~Wess,
  ``Extended supersymmetry and gauge theories,''
Nucl.\ Phys.\  B {\bf 133}, 275 (1978).
  %%CITATION = NUPHA,B133,275;%%  
  

\bibitem{KLRT-M}
S.~M.~Kuzenko, U.~Lindstr\"om, M.~Ro\v cek and G.~Tartaglino-Mazzucchelli,
``4D N=2 supergravity and projective superspace,'' 
JHEP {\bf 0809}, 051 (2008) [arXiv:0805.4683].

  
  
%\cite{Mezincescu:1979af}
\bibitem{Mezincescu}
L.~Mezincescu,
 ``On the superfield formulation of O(2) supersymmetry,''
Dubna preprint JINR-P2-12572 (June, 1979).
  %%CITATION = JINR-P2-12572;%%

%\cite{Howe:1981qj}
\bibitem{HST}
P.~S.~Howe, K.~S.~Stelle and P.~K.~Townsend,
``Supercurrents,''  Nucl.\ Phys.\  B {\bf 192}, 332 (1981).
  %%CITATION = NUPHA,B192,332;%%



%\cite{Butter:2010jm}
\bibitem{ButterK}
D.~Butter and S.~M.~Kuzenko,
``New higher-derivative couplings in 4D N = 2 supergravity,''
JHEP {\bf 1103}, 047 (2011) [arXiv:1012.5153 [hep-th]].

\bibitem{ButterK2}
D.~Butter and S.~M.~Kuzenko,
``N=2 AdS supergravity and supercurrents,''
JHEP \textbf{07}, 081 (2011)
%doi:10.1007/JHEP07(2011)081
[arXiv:1104.2153 [hep-th]].

\bibitem{Kuzenko:2008qz}
S.~M.~Kuzenko,
``On N = 2 supergravity and projective superspace: Dual formulations,''
Nucl. Phys. B \textbf{810}, 135 (2009)
%doi:10.1016/j.nuclphysb.2008.10.021
[arXiv:0807.3381 [hep-th]].


\bibitem{BKT}
I.~L.~Buchbinder, S.~M.~Kuzenko and A.~A.~Tseytlin,
``On low-energy effective actions in N = 2,4 superconformal 
theories  in  four dimensions,''
Phys.\ Rev.\ D {\bf 62},  045001 (2000) 
[arXiv:hep-th/9911221].


%\cite{Schwimmer:2010za}
\bibitem{SchT} 
A.~Schwimmer and S.~Theisen,
``Spontaneous breaking of conformal invariance and trace anomaly matching,''
Nucl.\ Phys.\ B {\bf 847}, 590 (2011)
[arXiv:1011.0696 [hep-th]].
  %%CITATION = ARXIV:1011.0696;%%

\bibitem{KST}
S.~M.~Kuzenko, A.~Schwimmer and S.~Theisen,
``Comments on anomalies in supersymmetric theories,''
J. Phys. A \textbf{53}, no.6, 064003 (2020)
%doi:10.1088/1751-8121/ab64a8
[arXiv:1909.07084 [hep-th]].

\bibitem{GHKSST}
J.~Gomis, P.~Hsin, Z.~Komargodski, A.~Schwimmer, N.~Seiberg and S.~Theisen,
``Anomalies, conformal manifolds, and spheres,''
JHEP \textbf{03}, 022 (2016)
%doi:10.1007/JHEP03(2016)022
[arXiv:1509.08511 [hep-th]].


\bibitem{deWGR}
B.~de Wit, M.~T.~Grisaru and M.~Ro\v{c}ek,
``Nonholomorphic corrections to the one-loop 
N=2 super Yang-Mills action,''
Phys.\ Lett.\ B {\bf 374},  297 (1996)
[arXiv:hep-th/9601115].


\bibitem{DeGiovanni:1997ib}
A.~De Giovanni, M.~T.~Grisaru, M.~Ro\v{c}ek, R.~von Unge and D.~Zanon,
``The N=2 superYang-Mills low-energy effective action at two loops,''
Phys. Lett. B \textbf{409}, 251 (1997)
%doi:10.1016/S0370-2693(97)00852-6
[arXiv:hep-th/9706013 [hep-th]].

\bibitem{BKO}
I.~L.~Buchbinder, S.~M.~Kuzenko and B.~A.~Ovrut,
``On the D = 4, N=2 non-renormalization theorem,''
Phys. Lett. B \textbf{433}, 335 (1998)
%doi:10.1016/S0370-2693(98)00688-1
[arXiv:hep-th/9710142 [hep-th]].


\bibitem{DKMSW}
N.~Dorey, V.~V.~Khoze, M.~P.~Mattis, M.~J.~Slater and W.~A.~Weir,
``Instantons, higher-derivative terms, and 
nonrenormalization theorems in
supersymmetric gauge theories,''
Phys.\ Lett.\ B {\bf 408}, 213 (1997) 
[arXiv:hep-th/9706007].


\bibitem{Matone}
D.~Bellisai, F.~Fucito, M.~Matone and G.~Travaglini,
``Non-holomorphic terms in N = 2 SUSY Wilsonian actions 
and RG equation,''
Phys.\ Rev.\ D {\bf 56}, 5218 (1997) 
[arXiv:hep-th/9706099].

\bibitem{DS}
M.~Dine and N.~Seiberg,
``Comments on higher derivative operators in some 
SUSY field theories,''
Phys.\ Lett.\ B {\bf 409},  239 (1997)
[arXiv:hep-th/9705057].



\bibitem{CT}
I.~Chepelev and A.~A.~Tseytlin,
 ``Long-distance interactions of branes: 
Correspondence between  supergravity
and super Yang-Mills descriptions,''
Nucl.\ Phys.\ B {\bf 515}, 73 (1998)
[arXiv:hep-th/9709087].

\bibitem{Tseytlin}
A.~A.~Tseytlin,
``Born-Infeld action, supersymmetry and string theory,''
in M. Shifman (Ed.), {\it The Many Faces of the Superworld}, 
World Scientific, 2000, pp. 417--452,
arXiv:hep-th/9908105.


\bibitem{Tseytlin96}
A.~A.~Tseytlin,
``Self-duality of Born-Infeld action and Dirichlet 3-brane 
of type IIB superstring theory,''
Nucl.\ Phys.\ B {\bf 469}, 51 (1996)
[arXiv:hep-th/9602064]; 

\bibitem{GG}
M.~B.~Green and M.~Gutperle,
``Comments on three-branes,''
Phys.\ Lett.\ B {\bf 377}, 28 (1996)
[arXiv:hep-th/9602077].

\bibitem{JKY}
A.~Jevicki, Y.~Kazama and T.~Yoneya,
``Quantum metamorphosis of conformal transformation 
in D3-brane  Yang-Mills theory,''
Phys.\ Rev.\ Lett.\  {\bf 81}, 5072 (1998) 
[arXiv:hep-th/9808039].

\bibitem{KM}
S.~M.~Kuzenko and I.~N.~McArthur,
``Quantum deformation of conformal symmetry 
in N = 4 super Yang-Mills  theory,''
Nucl.\ Phys.\ B {\bf 640}, 78 (2002) [arXiv:hep-th/0203236].

\bibitem{KMT}
S.~M.~Kuzenko, I.~N.~McArthur and S.~Theisen,
 ``Low energy dynamics from deformed conformal symmetry 
in quantum 4D N = 2 SCFTs,''
Nucl.\ Phys.\ B {\bf 660}, 131 (2003) [arXiv:hep-th/0210007].


\bibitem{Osborn2} 
  H.~Osborn,
  ``Local couplings and Sl(2,R) invariance for gauge theories at one loop,''
  Phys.\ Lett.\ B {\bf 561}, 174 (2003)
  %doi:10.1016/S0370-2693(03)00385-X
  [hep-th/0302119].


%\cite{Fradkin:1982xc}
\bibitem{FT1982} 
  E.~S.~Fradkin and A.~A.~Tseytlin,
  ``Asymptotic freedom in extended conformal supergravities,''
  Phys.\ Lett.\ B {\bf 110}, 117 (1982);
  %%CITATION = PHLTA,B110,117;%%
  ``One-loop beta function in conformal supergravities,''
  Nucl.\ Phys.\ B {\bf 203}, 157 (1982).
  %%CITATION = NUPHA,B203,157;%%

  
\bibitem{Paneitz}
  S.~M.~Paneitz,
 ``A quartic conformally covariant differential operator for 
 arbitrary pseudo-Riemannian manifolds,'' MIT preprint, March 1983; 
 published posthumously in:  SIGMA {\bf 4},  036 (2008)
  [arXiv:0803.4331 [math.DG]].

%\cite{Riegert:1984kt}
\bibitem{Riegert}
  R.~J.~Riegert,
  ``A non-local action for the trace anomaly,''
  Phys.\ Lett.\ B {\bf 134}, 56 (1984).
  %%CITATION = PHLTA,B134,56;%%


 \bibitem{BMZ} 
 D.~Brace, B.~Morariu and B.~Zumino, ``Duality invariant Born-Infeld theory,''
in {\it The Many Faces of the Superworld: Yury Golfand
Memorial Volume}, M. Shifman (Ed.), World Scientific, 2000, pp. 103--110 [hep-th/9905218].



\bibitem{K2020}
S.~M.~Kuzenko,
``Non-compact duality, super-Weyl invariance and effective actions,''
JHEP \textbf{07}, 222 (2020)
%doi:10.1007/JHEP07(2020)222
[arXiv:2006.00966 [hep-th]].


\bibitem{GWZ} 
  R.~Grimm, J.~Wess and B.~Zumino,
  ``Consistency checks on the superspace formulation of supergravity,''
  Phys.\ Lett.\ B {\bf 73}, 415 (1978);
  ``A complete solution of the Bianchi identities in superspace,''
 Nucl.\ Phys.\ B {\bf 152}, 255 (1979).


\bibitem{HT}
P.~S.~Howe and R.~W.~Tucker,
``Scale invariance in superspace,''
Phys.\ Lett.\ B {\bf 80}, 138 (1978).


\bibitem{Muller} M. M\"uller, {\it Consistent Classical Supergravity Theories},
(Lecture Notes in Physics, Vol. 336), Springer, Berlin, 1989. 

%\cite{Kuzenko:2008ry}
\bibitem{KT-M}
S.~M.~Kuzenko and G.~Tartaglino-Mazzucchelli,
``Different representations for the action principle in 4D N = 2 supergravity,''
 JHEP {\bf 0904}, 007 (2009) [arXiv:0812.3464 [hep-th]].
  %%CITATION = JHEPA,0904,007;%%

   
   
\end{thebibliography}
\end{document}